\documentclass[prc,twocolumn,showpacs,amsmath,amssymb]{revtex4}
\usepackage{graphicx}
\usepackage{dcolumn}
\usepackage{bm}
\usepackage{amstext}

\begin{document}

\title{Simple effective interaction: Infinite nuclear matter and finite nuclei}
\author{B. Behera$^1$, X. Vi\~nas$^2$, M. Bhuyan$^{1,3}$, T. R. Routray$^1$, 
B. K. Sharma$^2$ and S. K. Patra$^3$}
\affiliation{
$^1$ School of Physics, Sambalpur University, Jyotivihar-768 019, India.\\
$^2$ Department d'Estructura i Constitutuents de Materia, University de Barcelona, 
Diagonal 645, E-08028, Barcelona, Spain. \\
$^3$ Institute of Physics, Sachivalaya Marg, Bhubaneswar-751 005, India.\\ 
}

\date{\today}

\begin{abstract}
The mean field properties and equation of state for asymmetric nuclear 
matter are studied by using a simple effective interaction which has 
a single finite range Gaussian term. The study of finite nuclei with 
this effective interaction is done by means of constructing a quasilocal 
energy density functional for which the single particle equations take 
the form of Skryme-Hartree-Fock equations. The predictions of binding 
energies and charge radii of spherical nuclei are found to be compatible 
with the results of standard models as well as experimental data. 

\end{abstract}


\maketitle

\section{Introduction}
Last three decades have seen a regular interest to explain 
consistently the properties of nuclear matter, finite nuclei and 
nuclear reactions (nucleon-nucleon, nucleon-nucleus and nucleus-nucleus) 
with an effective interaction that has the efficiency to describe the two 
body system accurately. In this context, the study of nuclear 
properties from finite nuclei to highly $isospin$ asymmetric nuclear matter 
in a given model is a promising area of current interest. Relativistic and 
non-relativistic microscopic models such as Dirac-Brueckner-Hartree-Fock (DBHF) 
\cite{hoffma01,hoffman98,samma10,samma20,dbhf1,dbhf2,dbhf3,ebhf04,ebhf05}, 
Brueckner-Hartree-Fock (BHF) \cite{bhf1,bhf2,bhf3,bhf4,Bomb91,Wu07,bal04} 
and variational calculations using realistic interaction \cite{akmal01,Wiringa} 
are considered to be standard references in the regime of nuclear matter (NM).  
The {\it ab initio} extension of these models to finite nuclei are still 
in a preliminary stage. However, an energy density functional based on the 
microscopic calculations of \cite{bal04} that reproduces accurately binding 
energies and charge radii of finite nuclei has been reported recently 
\cite{baldo08}. Mean field models using effective interactions basically 
adopt the strategy of fitting the force parameters simultaneously to finite 
nuclei data and NM constraints (binding energy per particle, saturation density, 
incompressibility, symmetry energy, etc.). 

Finite nuclei results provided by Relativistic Mean Field (RMF) approaches 
are quite successful 
\cite{boguta77,sero86,ring90,patra01} in order to describe the $\beta$-stable 
and drip-line regions including superheavy elements. Some of these successful 
RMF interactions are NL3 \cite{lala97}, NL3* \cite{nl3im}, DD-ME1 \cite{ddme1}, 
DD-ME2 \cite{ddme2}, DD-F \cite{klahn06} and DD-ME$\delta$ \cite{roca11}.
In the non-relativistic frame the Skyrme \cite{brink72,stone12}, Gogny 
\cite{gogny80,gogny84,blaizot95,chappert08,goriely09} and 
M3Y \cite{hoffman98,nakada03,nakada08} effective interactions 
predict many properties of finite nuclei reasonably well. An important 
property of finite nuclei and infinite nuclear systems is the neutron and 
proton effective mass. However, mean field models predict neutron-proton 
{\it (n-p)} effective mass splittings in $isospin$ asymetric nuclear matter 
(ANM) not always in agreement with the results obtained in microscopic 
calculations. For instance, RMF models predict that the proton 
effective mass $m^*_p$ is larger than the neutron one $m^*_n$ 
\cite{kubis97,greco01,greco1a,greco03}, which is contrary to the results 
of microscopic calculations. In the non-relativistic frame, the Skyrme 
(with the exception of the {\it SLy} sets \cite{chab97,chab98,chab98a}) 
as well as the Gogny forces \cite{gogny80,gogny84,blaizot95,chappert08,
goriely09}, predict $n-p$ effective mass splittings which are, in general, 
consistent with the microscopic result  $m_n^* \textgreater m_p^*$.
Another important property of the mean field models is the mean field 
generated by the effective interaction, which in general is density and 
momentum dependent. In the non-relativistic frame the mean fields 
obtained by means of the Skyrme and finite range effective forces differ 
widely among them. The nuclear mean field corresponding to Skyrme forces 
has a $k^2$ dependence that is not in agreement with the behaviour of 
the mean field extracted from the analysis of the flow data in heavy-ion 
collision ({\t HIC}) experiments at intermediate and high energies 
\cite{bers,welke,gale,csernai,pan,zhang,Daneilewwicz00,fuchs06}. 
However, the more involved momentum dependence of the mean field provided 
by the finite range forces qualitatively agree with the behaviour extracted 
from the experiment.

In our earlier works \cite{trr98,trr02}, it has been shown that a mean 
field behaviour consistent with the experimental results can be obtained 
with finite range interactions containing a single form factor of 
{\it Yukawa}, {\it Gaussian} or {\it exponential} type. It was also 
shown \cite{trr98} that the range of these form factor and the strength of 
the exchange energy in symmetric nuclear matter (SNM)
could be constrained using the momentum dependence of the experimental optical 
potential extracted from nucleon-nucleus collision data \cite{welke,csernai}. 
In this way the different mean fields obtained with the aforementioned form 
factors have a similar behaviour over a wide range of momentum and density. 
This  simple effective interaction (SEI) has been used in several studies 
of NM, such as, momentum dependence of the mean field and  the equation of 
state (EOS) in ANM \cite{trr05,behera05}, neutron star matter at zero and 
finite temperature \cite{trr07,trr09} and the thermal evolution of NM 
properties \cite{trr11}. 

 The main aim of the present work is to show that this SEI, able to  
give an overall good description of NM properties in the isospin channel, 
can also reproduce 
finite nuclei properties with a similar quality to that obtained using the more
traditional effective interactions such as the Skyrme, Gogny and {\it M3Y}  
forces as well as with RMF parametrizations. Our study of ANM with the SEI,  
determines only nine out of the total eleven parameters of the interaction.
To determine the two remaining parameters is a necessary task for a wider 
application of the interaction. In this work we have adopted the procedure 
of fixing them from finite nuclei experimental data. Moreover, 
the knowledge of all the $11$ parameters of the SEI allows to make additional
NM calculations taking explicitly into account the $spin$- and 
$spin-isospin$ asymmetries. 

It is worth mentioning here that full Hartree-Fock (HF) calculations in 
finite nuclei with finite range effective interactions were performed 
first using the Gogny force\cite{gogny80,gogny84,blaizot95} in a harmonic 
oscillator basis. Finite nuclei HF calculations \cite{hoffman98} with the 
3-Yukawa (M3Y) force \cite{bert77} were performed first using the density matrix 
(DM) expansion \cite{negele72} in the limit of the local Fermi momentum 
approximation \cite{campi} which reduces the HF equations to a local form. 
Later on, both, full HF and  Hartree-Fock-Bogoliubov 
(HFB) calculations have been performed with the M3Y-interaction \cite{nakada03,nakada08} 
in NM and finite nuclei. An alternative approximation to the HF theory,
which is widely applied to electron systems, is based in the Kohn-Sham (KS) 
scheme 
\cite{kohn65} within the framework of the Density Functional Theory (DFT).
 We are aware that the application of the KS-DFT theory to self bound system 
like
nuclei is not obvious. However, it has been recently shown \cite{messud09}
that the {\it Hohenberg-Kohn} theorem \cite{hohen64}, on which the KS-DFT
scheme is based, has to be reformulated for the intrinsic frame in the case
of nuclei or other self quantum systems.
The original version of the KS-DFT theory is local,
however, in the nuclear case, non-local contributions such as the effective mass
and the spin-orbit potential are essential ingredients of the energy density
owing to the momentum dependence of the nuclear interaction.
The non-local extension of the DFT was treated first by Gilbert \cite{gilbert75}.
Later on, a modification of the non-local generalization of the DFT and its 
quasilocal reduction were discussed in detail in Ref.~\cite{vinas03} and we
refer the reader to this reference for more details. In order to  write 
explicitly
the exchange energy in a local form, it is necessary to express the DM in terms
of only local quantities, such as the particle and kinetic energy densities.
This can be done, for instance, using the DM expansion of Negele and Vautherin 
\cite{negele72} or Campi and Bouyssy \cite{campi}, or, alternatively, with the 
semiclassical $\hbar$-expansion of the DM \cite{vinas00}. 
In this work we will use this latter approximation together with the SEI to 
build up an energy density functional able to describe ground state properties 
of finite nuclei. It is also important to point out that nowadays a modified DM 
expansion is used in Effective Field Theory calculations with realistic interactions 
to write the microscopic energy density in a local Skyrme-like form
(see \cite{stoit10} and references therein).

The paper is organized as follows. In Section II we present the theoretical 
formalism for ANM with the SEI containing a single Gaussian form factor 
for describing the finite range part. 
The parameters necessary for a complete description 
of SNM and ANM are discussed together with the results. In section III, 
a general formulation of NM having all the three possible asymmetries, 
namely isospin, spin and spin-isospin, is given in order to examine the 
behaviour of the SEI in the spin and spin-isospin channels. In Section IV we 
build up an energy density functional based on the SEI within the framework of the 
quasilocal 
DFT. In the same Section the predictive power of the SEI to describe the ground state 
binding energies and charge radii of spherical nuclei is analyzed with some detail.
Finally, Section IV contains a brief summary and conclusions.

\section{Isospin Asymmetric Infinite Nuclear Matter}

Our proposed SEI, that we will use in the calculation of  
NM and finite nuclei properties, has the following form  
\begin{widetext}
\begin{eqnarray}
v_{eff}(r)&=&t_0 (1+x_0P_{\sigma})\delta(r) 
+\frac{t_3}{6}(1+x_3 P_{\sigma})\left(\frac{\rho({\bf R})}
{1+b\rho({\bf R})}\right)^{\gamma} \delta(r) 
+ \left(W+BP_{\sigma}-HP_{\tau}-MP_{\sigma}P_{\tau}\right)f(r),
\label{eq1}
\end{eqnarray} 
where, $f(r)$ is the form factor of the finite range interaction 
which depends on a single parameter $\alpha$, the range of the interaction. 
In this work $f(r)$ is 
chosen to be of  Gaussian form, $e^{-r^2/\alpha^2}$. The other terms 
have their usual meaning \cite{trr98,trr02}. The modified density 
dependence with the parameter $b$ in the denominator is taken, as in 
the earlier works \cite{trr05,trr07,trr09,trr11}, in order to prevent 
the supraluminous behaviour in NM. The SEI in Eqn. (1) contains 
$11$-parameters, namely $t_0$, $x_0$, $t_3$, $x_3$, $b$, $W$, $B$, 
$H$, $M$, $\gamma$ and $\alpha$ that will be determined from NM and finite 
nuclei properties. 

The energy density functional $H_{T}(\rho_n, \rho_p)$ in ANM at temperature 
$T$, derived using the SEI in Eqn. (\ref{eq1}), can be expressed in terms of the 
neutron (proton) densities $\rho_n$ ($\rho_p$) and their respective momentum 
distribution functions $f_{T}^n({\bf k})$ ($f_{T}^p({\bf k})$) as \cite{trr09}:

\begin{eqnarray}
H_T(\rho_n, \rho_p) &=& \frac{\hslash^2}{2m}\int\left[ f_{T}^n({\bf k})
+f_{T}^p({\bf k})\right]k^2d^3k \nonumber \\
&& + \frac{1}{2}\left[\frac{\varepsilon_0^l}{\rho_0}+\frac{\varepsilon_{\gamma}^l}
{\rho_0^{\gamma+1}}\left(\frac{\rho}{1+b\rho}\right)^{\gamma}\right]\left(\rho_n^2+\rho_p^2\right)
+\left[\frac{\varepsilon_0^{ul}}{\rho_0}+\frac{\varepsilon_{\gamma}^{ul}}
{\rho_0^{\gamma+1}}\left(\frac{\rho}{1+b\rho}\right)^{\gamma}\right]\rho_n\rho_p \nonumber \\
&& +\frac{\varepsilon_{ex}^l}{2\rho_0}\int\int\left[f_T^n({\bf k})f_{T}^n
({\bf k'})+f_T^p({\bf k})f_{T}^p({\bf k'})\right]
g_{ex}\left(\vert{\bf k}-{\bf k'}\vert\right)d^3k d^3k' \nonumber \\
&& +\frac{\varepsilon_{ex}^{ul}}{2\rho_0}\int\int\left[f_T^n({\bf k})f_{T}^p
({\bf k'})+f_T^p({\bf k})f_{T}^n({\bf k'})\right]
g_{ex}\left(\vert{\bf k}-{\bf k'}\vert\right)d^3k d^3k',
\label{eq2}
\end{eqnarray}
where, $\rho_0$ is the saturation density of SNM and  $g_{ex}({\bf k})=
\frac{\int e^{i{\bf k}\cdot{\bf r}}f(r)d^3r}{\int f(r)d^3r} $ is the 
normalized Fourier transform of the finite range form factor $f({r})$. 
The index $l$ ($ul$) is used for the interaction between a pair of like 
(unlike) nucleons. For the sake of simplicity, we have assumed that the 
interaction between a like and unlike pair of nucleons have the same range 
but differ in strengths. The energy density in Eqn. (\ref{eq2}) is written in terms of
nine parameters, namely, $\gamma$, b, $\varepsilon_{0}^{l}$, $\varepsilon_{0}^{ul}$, 
$\varepsilon_{\gamma}^{l}$,$\varepsilon_{\gamma}^{ul}$, $\varepsilon_{ex}^{l}$, 
$\varepsilon_{ex}^{ul}$ and $\alpha$. The six new parameters are connected to 
the interaction parameters of Eq.(\ref{eq1})
through the relations  
\begin{eqnarray}
\varepsilon_{0}^{l}=\rho_0\left[\frac{t_0}{2}\left(1-x_0\right)
+\left(W+\frac{B}{2}-H-\frac{M}{2}\right)\int f(r)d^3r\right] \nonumber \\
\nonumber \\
\varepsilon_{0}^{ul}=\rho_0\left[\frac{t_0}{2}\left(2+x_0\right)
+\left(W+\frac{B}{2}\right)\int f(r)d^3r\right] \nonumber \\
\nonumber \\
\varepsilon_{\gamma}^{l}=\frac{t_3}{12}\rho_0^{\gamma+1}(1-x_3),      
\varepsilon_{\gamma}^{ul}=\frac{t_3}{12}\rho_0^{\gamma+1}(2+x_3) \nonumber \\
\nonumber \\
\varepsilon_{ex}^{l}=\rho_0\left(M-\frac{W}{2}-B+\frac{H}{2}\right)
\int f(r)d^3r \nonumber \\
\nonumber \\
\varepsilon_{ex}^{ul}=\rho_0\left(M+\frac{H}{2}\right)
\int f(r)d^3r.
\end{eqnarray}
The neutron (proton) single particle energy 
$\epsilon_T^{n(p)}({\bf k},\rho_n,\rho_p$) can be obtained as the 
functional derivative of the energy density 
$H_T(\rho_n,\rho_p)$ in Eqn.(\ref{eq2}) with respect to the neutron (proton) 
and is given by 
\begin{eqnarray}
\epsilon_T^{n(p)}({\bf k},\rho_n,\rho_p)=\frac{\hslash^2k^2}{2m}
+u_T^{n(p)}({\bf k},\rho_n,\rho_p),
\end{eqnarray}
where, the first term corresponds to the kinetic energy of neutrons (protons) 
and $u_T^{n(p)}$ is the corresponding mean fields. 

The Fermi-Dirac momentum distribution functions $f_T^n({\bf k})$ and 
$f_T^p({\bf k})$ at finite $T$ take the form of step-functions in the $T$=0 
limit. In this case the mean fields $u^{n(p)}({\bf k},\rho_n,\rho_p)$ and 
energy density $H(\rho_n,\rho_p)$ can be evaluated analytically. The 
corresponding expressions for a Gaussian form factor $f(r)$ read, 
\begin{eqnarray}
H(\rho_n,\rho_p)&=&\frac{3\hslash^2}{10m}\left(k_n^2\rho_n+k_p^2\rho_p\right)
+\frac{\varepsilon_{0}^{l}}{2\rho_0}\left(\rho_n^2+\rho_p^2\right)
+\frac{\varepsilon_{0}^{ul}}{\rho_0}\rho_n\rho_p \nonumber \\
&&+\left[\frac{\varepsilon_{\gamma}^{l}}{2\rho_0^{\gamma+1}}\left(\rho_n^2+\rho_p^2\right)
+\frac{\varepsilon_{\gamma}^{ul}}{\rho_0^{\gamma+1}}\rho_n\rho_p\right]
\left(\frac{\rho({\bf R})}{1+b\rho({\bf R})}\right)^{\gamma} \nonumber \\
&&+\frac{\varepsilon_{ex}^{l}}{2\rho_0}\rho_n^2\left[\frac{3\Lambda^6}{16k_n^6}
-\frac{9\Lambda^4}{8k_n^4}+\left(\frac{3\Lambda^4}{8k_n^4}-\frac{3\Lambda^6}{16k_n^6}\right)
e^{-4k_n^2/\Lambda^2}+\frac{3\Lambda^3}{2k_n^3}\int_0^{2k_n/\Lambda}e^{-t^2}dt\right] \nonumber \\
&&+\frac{\varepsilon_{ex}^{l}}{2\rho_0}\rho_p^2\left[\frac{3\Lambda^6}{16k_p^6}
-\frac{9\Lambda^4}{8k_p^4}+\left(\frac{3\Lambda^4}{8k_p^4}-\frac{3\Lambda^6}{16k_p^6}\right)
e^{-4k_p^2/\Lambda^2}+\frac{3\Lambda^3}{2k_p^3}\int_0^{2k_p/\Lambda}e^{-t^2}dt\right] \nonumber \\
&&+\frac{\varepsilon_{ex}^{ul}\rho_n}{\rho_0}\frac{1}{\Lambda^2}\int_0^{k_p}dkk^2
\left[\frac{3\Lambda^4}{8kk_n^3}\left\lbrace e^{-\left(\frac{k+k_n}{\Lambda}\right)^2}
-e^{-\left(\frac{k-k_n}{\Lambda}\right)^2}\right\rbrace
+\frac{3\Lambda^3}{4k_n^3}\int_{\left(\frac{k-k_n}{\Lambda}\right)}^{\left(\frac{k+k_n}{\Lambda}
\right)}e^{-t^2}dt\right],
\end{eqnarray}

and

\begin{eqnarray}
u^{n(p)}({\bf k},\rho_n,\rho_p)&=& \frac{\varepsilon_{0}^{l}\rho_{n(p)}}{\rho_0}
+\frac{\varepsilon_{0}^{ul}\rho_{p(n)}}{\rho_0}
+\left(\frac{\varepsilon_{\gamma}^{l}\rho_{n(p)}}{\rho_0^{\gamma+1}}
+\frac{\varepsilon_{\gamma}^{ul}\rho_{p(n)}}{\rho_0^{\gamma+1}}\right) 
\left(\frac{\rho({\bf R})}{1+b\rho({\bf R})}\right)^{\gamma} \nonumber \\
&&+\frac{\varepsilon_{ex}^{l}\rho_{n(p)}}{\rho_0}\left[\frac{3\Lambda^4}{8kk_{n(p)}^3}
\left\lbrace e^{-\left(\frac{k+k_{n(p)}}{\Lambda}\right)^2}
-e^{-\left(\frac{k-k_{n(p)}}{\Lambda}\right)^2}\right\rbrace
+\frac{3\Lambda^3}{4k_{n(p)}^3}\int_{\left(\frac{k-k_{n(p)}}{\Lambda}\right)}
^{\left(\frac{k+k_{n(p)}}{\Lambda}\right)}e^{-t^2}dt\right] \nonumber \\
&&+\frac{\varepsilon_{ex}^{ul}\rho_{p(n)}}{\rho_0}\left[\frac{3\Lambda^4}{8kk_{p(n)}^3}
\left\lbrace e^{-\left(\frac{k+k_{p(n)}}{\Lambda}\right)^2}
-e^{-\left(\frac{k-k_{p(n)}}{\Lambda}\right)^2}\right\rbrace
+\frac{3\Lambda^3}{4k_{p(n)}^3}\int_{\left(\frac{k-k_{p(n)}}{\Lambda}\right)}
^{\left(\frac{k+k_{p(n)}}{\Lambda}\right)}e^{-t^2}dt\right] \nonumber \\
&&+\left\lbrace\frac{\varepsilon_{\gamma}^{l}}{2\rho_0^{\gamma+1}}\left(\rho_n^2+\rho_p^2\right)
+\frac{\varepsilon_{\gamma}^{ul}}{2\rho_0^{\gamma+1}}\rho_n\rho_p\right\rbrace
\frac{\gamma\rho^{\gamma-1}}{(1+b\rho)^{\gamma+1}}.
\label{eq6}
\end{eqnarray}
Here $\rho=\rho_n+\rho_p$ is the total nucleonic density, 
$k_{n(p)}=\left(3\pi^2\rho_{n(p)}\right)^{\frac{1}{3}}$ is the neutron 
(proton) Fermi momentum and $\Lambda=2/\alpha$. 
The explicit expression for the energy density and other relevant quantities
in ANM obtained with a Yukawa form factor can be found in Refs.
\cite{trr05,trr07,trr09,trr11}. It is worth mentioning here that compact
formulae for energy in NM for several finite range form factors can be found 
in Appendix A of Ref.~\cite{nakada03}. 

In the limit of symmetric nuclear matter, $\rho_n=\rho_p=\rho/2$, the 
corresponding expressions $H(\rho)$ and $u({\bf k},\rho)$ become, 
\begin{eqnarray}
H(\rho) &=& \rho e(\rho)=\frac{3\hslash^2k_f^2\rho}{10m}
+\frac{(\varepsilon_{0}^{l}+\varepsilon_{0}^{ul})}{4\rho_0}\rho^2
+\frac{(\varepsilon_{\gamma}^{l}+\varepsilon_{\gamma}^{ul})}{4\rho_0^{\gamma+1}}
\rho^2\left(\frac{\rho({\bf R})}
{1+b\rho({\bf R})}\right)^{\gamma} \nonumber \\
 &&+\frac{(\varepsilon_{ex}^{l}+\varepsilon_{ex}^{ul})}{4\rho_0}
\rho^2\left[\frac{3\Lambda^6}{16k_n^6}-\frac{9\Lambda^4}{8k_n^4}
+\left(\frac{3\Lambda^4}{8k_n^4}-\frac{3\Lambda^6}{16k_n^6}\right)
e^{-4k_n^2/\Lambda^2}+\frac{3\Lambda^3}{2k_n^3}\int_0^{2k_n/\Lambda}e^{-t^2}dt\right] 
\label{eq7}
\end{eqnarray}
and
\begin{eqnarray}
u({\bf k},\rho)&=&\frac{(\varepsilon_{0}^{l}+\varepsilon_{0}^{ul})}{2\rho_0}\rho
+\frac{(\varepsilon_{\gamma}^{l}+\varepsilon_{\gamma}^{ul})}{2\rho_0^{\gamma+1}}
\left(\frac{\rho({\bf R})}{1+b\rho({\bf R})}\right)^{\gamma} 
\left(1+b\rho+\frac{\gamma}{2}\right)\nonumber \\
&&+\frac{(\varepsilon_{\gamma}^{l}+\varepsilon_{ex}^{ul})}{2\rho_0}\rho
\left[\frac{3\Lambda^4}{8kk_f^3}
\left\lbrace e^{-\left(\frac{k+k_f}{\Lambda}\right)^2}
-e^{-\left(\frac{k-k_f}{\Lambda}\right)^2}\right\rbrace
+\frac{3\Lambda^3}{4k_f^3}\int_{\left(\frac{k-k_f}{\Lambda}\right)}
^{\left(\frac{k+k_f}{\Lambda}\right)}e^{-t^2}dt\right],
\label{eq8}
\end{eqnarray}
where $k_f=\left(\frac{3\pi^2}{2}\rho\right)^{1/3}$ is the Fermi momentum 
and $e(\rho)$ is the energy per particle in SNM. The new parameter 
combinations,
\begin{eqnarray}
\left(\frac{\varepsilon_{0}^{l}+\varepsilon_{0}^{ul}}{2}\right)=\varepsilon_0,    
\left(\frac{\varepsilon_{\gamma}^{l}+\varepsilon_{\gamma}^{ul}}{2}\right)=\varepsilon_{\gamma},   
\left(\frac{\varepsilon_{ex}^{l}+\varepsilon_{ex}^{ul}}{2}\right)=\varepsilon_{ex},
\label{eq9}
\end{eqnarray} 
\end{widetext} 
alongwith $\gamma$, $b$ and $\alpha$ are the six parameters needed 
for a complete description of the mean field properties and the EOS 
in SNM. The way of determining the parameters which enter in 
Eqs.~(\ref{eq7}) and (\ref{eq8}) in this case is similar to the one 
described in Ref.~\cite{trr07} for a {\it Yukawa} form factor. The only 
difference in this work is that we take  
the kinetic energy term in its non-relativistic form for convenience of
application to finite nuclei. The range parameter $\alpha$ and the  
exchange strength combination $\varepsilon_{ex}$ in SNM are 
determined by adopting a simultaneous optimization procedure with the 
constraint that the attractive optical potential changes sign for a kinetic 
energy $\frac{\hslash^2k^2}{2m}$=300 MeV of the incident nucleon
(see Ref.~\cite{trr98} for details). In this way the
values $\varepsilon_{ex}$=-94.46 MeV and $\alpha$=0.7596 $fm$ are found  
where we have used only the standard values of the 
nucleon mass $m$=939 MeV, Fermi kinetic energy $\hbar^2 k^2_{f_0}/2m$=36.4 MeV
(the Fermi momentum is $k_{f_0}=(3 \pi^2 \rho_0/2)^{1/3}$ with a 
saturation density $\rho_0$=0.157 $fm^{-3}$) and energy per particle 
in SNM $e(\rho_0)$= -16.0 MeV. 
The momentum dependence of the nuclear mean field in SNM computed 
using this SEI for three different densities, $\rho$=0.1, 0.3, and 
0.5 $fm^{-3}$, is displayed in the upper panel of the Fig. 1. In order 
to compare with the predictions 
of the realistic interaction UV14+UVII \cite{Wiringa}, we plot 
$u^{ex}({\bf k},\rho)=\left[u({\bf k},\rho)-u({\bf k_f},\rho)\right]$
as functions of $k$. Here the saturation properties of the
interaction have been substracted because they may be model dependent.
We find a quite good agreement between the results computed with the SEI 
and the microscopic predictions over a wide range of momenta and densities.
The same comparison is shown in the lower panel of Fig 1. for Gogny 
D1 \cite{gogny80}, D1S \cite{blaizot95} and D1M \cite{goriely09} sets with 
the microscopic results \cite{Wiringa}. We see that the momentum dependence 
of the Gogny D1 force agrees well with the microscopic results upto density 
$\rho=0.3$ $fm^{-3}$ over the whole range of momentum shown in the figure, 
but agreement of the results become qualitative at higher density as can be 
seen from the comparison for the curve corresponding to $\rho=0.5$ $\it fm^{-3}$. 
On the otherhand, the Gogny D1S and D1M forces do not follow the trends of the 
microscopic calculation beyond $k$ > 2 $fm^{-1}$.
It may be pointed out that the Gogny forces are fitted basically 
to finite nuclei data and, therefore, there is no reason {\it a priori} for 
reproducing the momentum dependence of the microscopic calculation in NM 
over a wide range of momentum in case of all the Gogny force sets.

The effective mass in SNM is given by, 
\begin{eqnarray}
\left[\frac{m^*}{m}(k,\rho)\right]=\left[1+\frac{m}{\hslash^2k}
\frac{\partial u(k,\rho)}{\partial k}\right]^{-1},
\end{eqnarray}
which is momentum and density dependent and can be 
calculated once the exchange part of the mean field in Eqn. (\ref{eq8}) 
is known. For the values of the exchange strength and range determined 
above, the prediction of the effective mass at saturation density 
$\rho=\rho_0$ and momentum $k=k_{f_0}$ is $m^*/m$=0.709. 
The parameter $b$ has been adjusted to avoid a supraluminous behaviour at high 
densities in SNM at $T$=0 (see Ref.~\cite{trr97} for more details). 
With the standard values mentioned before we obtain $b$=0.5914 $fm^3$.
The two remaining strength parameters $\varepsilon_{0}$ and 
$\varepsilon_{\gamma}$ are obtained from the saturation conditions. 
The exponent $\gamma$, that determines the stiffness of the EOS of SNM, is 
taken as $\gamma$=$\frac{1}{2}$ which gives a value of incompressibility 
at normal NM density, $K(\rho_0)$=245 MeV. 
With the six parameters of SNM 
determined, one can calculate the energy per particle $e (\rho)$ and pressure 
$P (\rho)$ in SNM. These quantities are displayed as a functions of the density 
in Fig. 2(a) and 2 (b), respectively. They are compared with the values of 
$e (\rho)$ and $P (\rho)$ predicted by some microscopic calculations 
and several non-relativistic and relativistic models. 
In Fig. 2(b) we also display the band of allowed values of $P(\rho)$
in the range $2- 4.6$ $fm^{-3}$ extracted from the analysis of the 
flow data in high energy HIC experiments \cite{Daneilewwicz02}.
Further experimental information of the pressure-density relationship 
is provided by the analysis of $K^+$ production data \cite{lynch09} in the 
low density domain between $1.2-2\rho_0$ $fm^{-3}$. We find that our 
SEI curve $P(\rho)$, obtained using a value of $\gamma$=1/2 passes well
within these experimentally extracted regions.

The complete study of ANM now requires the splitting of the three 
strength parameters
$\varepsilon_{ex}=\frac{(\varepsilon_{ex}^l+\varepsilon_{ex}^{ul})}{2}$,
$\varepsilon_{0}=\frac{(\varepsilon_{0}^l+\varepsilon_{0}^{ul})}{2}$ and
$\varepsilon_{\gamma}=\frac{(\varepsilon_{\gamma}^l+\varepsilon_{\gamma}^{ul})}{2}$
into two specific channels for interactions between like 
and unlike pairs of nucleons. 
We have fixed the splitting of $\varepsilon_{ex}$
into the like channel to be $\varepsilon_{ex}^l=2\varepsilon_{ex}/3$.
This is the critical value for which the thermal evolution of NM properties
in pure neutron matter (PNM) does not surpass the SNM results at any density
for any temperature \cite{trr11}.
The splitting of the exchange strength parameter $\varepsilon_{ex}$ 
into $(l)$ and $(ul)$ channels fully determines the $n,p$-
effective mass behaviour in ANM. The $n,p$- effective masses
in ANM are calculated starting from their usual definitions,
\begin{eqnarray}
\left[\frac{m^*}{m}(k,\rho_n,\rho_p)\right]_{n,p}=\left[1+\frac{m}{\hslash^2k}
\frac{\partial u^{n,p}(k,\rho_n,\rho_p)}{\partial k}\right]^{-1}.
\label{eq11}
\end{eqnarray}
The magnitude of the splitting of the n,p-effective masses in ANM at 
normal density, $[m^*/m]_n-[m^*/m]_p$, obtained with 
the SEI is shown in Fig. 3 as a function of the $isospin$ asymmetry 
$\beta=\frac{(\rho_n-\rho_p)}{\rho}$ alongwith the results of microscopic 
calculations and the values obtained with the Gogny effective forces. 
Our result compares well with the prediction of the DBHF calculations using the 
Bonn potential \cite{samma10} over the whole range of asymmetry $\beta$. The 
Gogny D1 interaction also predict closely the similar trend as that of our SEI. 
But, the $n-p$ effective mass splittings obtained using the Gogny D1S and D1M 
interactions are very similar between them and have a smaller value as compared 
with DBHF result. On the other hand, the splittings in the cases of BHF+3-BF 
and EBHF+3-BF \cite{ebhf04,ebhf05,bhf21} calculations are found to be close 
to each other, predicting a larger $n-p$ effective mass splitting than the DBHF 
model. 

Once the like and unlike components of $\varepsilon_{ex}$ are fixed, 
the splitting of the two remaining strength parameters combinations, namely
$\varepsilon_{0}=\frac{(\varepsilon_{0}^l+\varepsilon_{0}^{ul})}{2}$ and
$\varepsilon_{\gamma}=\frac{(\varepsilon_{\gamma}^l+\varepsilon_{\gamma}^{ul})}{2}$
can be obtained from values of symmetry energy $E_s(\rho_0)$ and
its derivatives $E'_s(\rho_0)=\rho_0\frac{dE_s(\rho)}{d\rho}$ computed at saturation
density. We have assumed a value of $E_s(\rho_0)$=35.0 MeV which is within the possible 
range between 30-35 MeV \cite{luo09,stone07,chen08,tsang12,stone12}. The value of
$E'_s(\rho_0)$ is ascertained from an universal high density behaviour 
of the asymmetric contribution of the nucleonic part of the energy density 
in charge neutral {\it beta}-stable $n+p+e+\nu$ matter \cite{trr07,klahn06}.
The value $E'_s(\rho_0)$ =25.42 MeV obtained in this way predicts a slope parameter 
(symmetry pressure) $L(\rho_0)$=76.3 MeV that is well within the range $70\pm15$ 
MeV of recent Finite Range Droplet Model (FRDM) prediction \cite{moller12}.
Our value $L(\rho_0)$ also lie within the window 45-80 MeV estimated from a 
compilation
of different predictions of $L$ deduced from experiments involving antiprotonic 
atoms, heavy-ion reactions, proton scattering, nuclear masses, microscopic 
calculations and giant dipole resonances \cite{warda09}. 
The values of the nine parameters altogether necessary for a complete 
description of SNM and ANM alongwith the NM properties (saturation density
$\rho_0$, 
binding energy per nucleon in SNM $e(\rho_0)$, incompressibility $K(\rho_0)$, 
symmetry energy $E_s(\rho_0)$ and slope parameter $L(\rho_0)$)  
are given in Table I. The strength parameters in SNM can be obtained from this 
table using the relations of Eqn. (\ref{eq9}).  
 
Once the nine parameters that fully determine the ANM are known, the energy per 
particle in PNM, $e^N(\rho)$, can be easily calculated and their values as a function 
of the density $\rho$ are displayed in Fig. 4 alongwith the results of realistic 
\cite{pandaripande81} and some effective models. In the low density region, 
$\rho<0.12\rho_0$, it has been verified that our curve 
passes well through the region predicted by different microscopic calculations 
performed using Quantum Monte Carlo techniques \cite{gezerlis10}
as it can be seen in the inset of this figure. 
In the high density region the SEI curve shows a stiff behaviour because of the 
relatively high value of $E_s(\rho)$. The effective M3Y-P5 \cite{nakada08} curve 
has a relatively more soft behaviour throughout the density range but maintains 
an increasing trend unlike the case with the D1S Gogny force displayed in the same figure.
Further, the EOS of PNM predicts a ratio $3P^N(\rho_0)/(L\rho_0)$ =1.0 MeV $fm^{-3}$, 
which is in agreement with the value obtained by Piekarewicz \cite{piekarewicz07} 
from an analysis of the relation between the slope parameter $L$ and the 
pressure of PNM at normal density $\rho_0$. 
The density dependence of nuclear symmetry energy, 
$E_s(\rho)$, calculated as the difference between the energy per particle in PNM 
and SNM, is displayed in Fig. 5. It compares well with the results provided by 
the realistic A18+$\delta$v+UIX* interaction \cite{akmal01} upto a density 
0.2 $fm^{-3}$ and follows a similar trend till a density 0.8 $fm^{-3}$. Our 
curve also compares reasonably well with the DBHF \cite{ebhf04,ebhf05} results 
upto a density of 0.5 $fm^{-3}$, but clearly differs beyond this point where this 
microscopic calculation shows a very stiff behaviour of the symmetry energy with
increasing density.   

\section{ nuclear matter under isospin, spin and spin-isospin asymmetries}

In the previous section we have restricted our NM study to the case of isospin 
asymmetric NM, i.e., ANM, which has much physical relevance from the 
point of view of the available experimental and empirical constraints. However, 
in order to examine the predictions of SEI in the spin and spin-isospin channels, 
we shall make first the explicit formulation of NM having both spin and isospin asymmetries. 
The total NM density is now given by 
\begin{eqnarray}
\rho=\rho_{nu}+\rho_{nd}+\rho_{pu}+\rho_{pd},
\end{eqnarray}
where, $nu$, $nd$ ($pu$, $pd$) denote the neutron spin-up and spin-down  
(proton spin-up and spin-down) states and $\rho_{nu}$, $\rho_{nd}$, 
$\rho_{pu}$ and $\rho_{pd}$ are the respective densities. The 
Fermi-momenta corresponding to these four densities are expressed as 
\begin{eqnarray}
k_{i,j} = \left(6\pi^2\rho_{i,j}\right)^{1/3},
\end{eqnarray}
with $i$=n,p and $j$=u,d. Making Taylor expansion of the energy density 
around the spin-saturated SNM value $H(\rho)$ and keeping terms upto lowest 
order only, one can obtain
\begin{eqnarray}
H(\rho_{nu},\rho_{nd},\rho_{pu},\rho_{pd}) =
H(\rho) +\frac{1}{2}\alpha_{\tau}^2\rho E_{\tau}(\rho) \nonumber \\ 
+\frac{1}{2}\alpha_{\sigma}^2\rho E_{\sigma}(\rho) 
+\frac{1}{2}\alpha_{\sigma\tau}^2 \rho E_{\sigma \tau}(\rho),
\end{eqnarray} 
where $\alpha_{\tau}$, $\alpha_{\sigma}$ and $\alpha_{\sigma\tau}$ are the $isospin$, 
$spin$ and $spin-isospin$ asymmetries defined as 
\begin{eqnarray}
\alpha_{\tau}=\frac{\left(\rho_{nu}+\rho_{nd}\right)-\left(\rho_{pu}+\rho_{pd}\right)}
{\rho} \nonumber \\
\alpha_{\sigma}=\frac{\left(\rho_{nu}+\rho_{pu}\right)-\left(\rho_{nd}+\rho_{pd}\right)}
{\rho} \nonumber \\
\alpha_{\sigma\tau}=\frac{\left(\rho_{nu}-\rho_{pu}\right)-\left(\rho_{nd}-\rho_{pd}\right)}
{\rho},
\end{eqnarray}
and $E_{\tau}$, $E_{\sigma}$ and $E_{\sigma \tau}$ are the respective symmetry 
energies. The expressions of these symmetry energies are given by, 
\begin{eqnarray}
E_{\tau} (\rho) = \frac{\hbar^2 k_f^2}{3m}-\frac{\rho}{2}
\int\left(\frac{3}{8}v^{te}-\frac{1}{8}v^{se}-\frac{3}{8}v^{to}
+\frac{1}{8}v^{so}\right)d^3r \nonumber \\
-\frac{\rho}{2} \int\left(\frac{3}{8}v^{te}-\frac{1}{8}v^{se}+\frac{3}{8}v^{to}
-\frac{1}{8}v^{so}\right)j_0^2 (x_f) d^3r \nonumber \\
-\rho \int\left(\frac{3}{16}v^{se}+\frac{3}{16}v^{te}-\frac{9}{16}v^{to}
-\frac{1}{16}v^{so}\right)j_1^2 (x_f) d^3r \nonumber \\
\nonumber \\
E_{\sigma} (\rho) = \frac{\hbar^2 k_f^2}{3m}-\frac{\rho}{2}
\int\left(\frac{3}{8}v^{se}-\frac{1}{8}v^{te}-\frac{3}{8}v^{to}
+\frac{1}{8}v^{so}\right)d^3r \nonumber \\
-\frac{\rho}{2} \int\left(\frac{3}{8}v^{se}-\frac{1}{8}v^{te}+\frac{3}{8}v^{to}
-\frac{1}{8}v^{so}\right)j_0^2 (x_f) d^3r \nonumber \\
-\rho \int\left(\frac{3}{16}v^{se}+\frac{3}{16}v^{te}-\frac{9}{16}v^{to}
-\frac{1}{16}v^{so}\right)j_1^2 (x_f) d^3r \nonumber \\
\nonumber \\
E_{\tau \sigma} (\rho) = \frac{\hbar^2 k_f^2}{3m}-\frac{\rho}{2}
\int\left(\frac{1}{8}v^{se}+\frac{1}{8}v^{te}-\frac{1}{8}v^{to}
-\frac{1}{8}v^{so}\right)d^3r \nonumber \\
-\frac{\rho}{2} \int\left(\frac{1}{8}v^{se}+\frac{1}{8}v^{te}+\frac{1}{8}v^{to}
+\frac{1}{8}v^{so}\right)j_0^2 (x_f) d^3r \nonumber \\
-\rho \int\left(\frac{3}{16}v^{se}+\frac{3}{16}v^{te}-\frac{9}{16}v^{to}
-\frac{1}{16}v^{so}\right)j_1^2 (x_f) d^3r, \nonumber \\
\label{eq16}
\end{eqnarray}
where, $j_l, l=0,1$ are the spherical Bessel function of order $l$, $x_f=\alpha k_f$, 
$k_f$ being the Fermi-momentum in $spin$ saturated SNM, and $\alpha$ is the range of the 
interaction. In these equations $v^{se}$, $v^{te}$, $v^{to}$ and $v^{so}$ are the 
interactions in the 
$singlet-even$, $triplet-even$, $triplet-odd$ and $singlet-odd$ states 
respectively. Calculation of the symmetry energies using Eqn. (\ref{eq16}) requires to 
know 
the interactions in all these four 
different states independently. In the case of the SEI, this is possible 
provided that all the eleven parameters of the interaction are known. The study of ANM 
performed in the previous section allows to fix nine of them. The two open parameters 
($t_0$ and $x_0$ considered here) are determined from the study of finite nuclei as will 
be explained in 
the next Section. To complet the study of NM matter in the spin and spin-isospin channels 
we will use the numerical value of the parameters of the SEI interaction reported in Table II. 

We first calculate the contributions of the 
$singlet-odd$ (SO), $singlet-even$ (SE), $triplet-odd$ (TO) and $triplet-even$ 
(TE) states to the potential energy per nucleon $\langle V \rangle /A$ in $spin$ 
saturated SNM. These contributions, computed using the SEI, are displayed 
in the four panels of Fig.6 as functions of the Fermi-momentum $k_f$ along with the 
results for 
the Gogny D1S and M3Y-P5 interactions. The odd-states contributions of SEI and 
D1S have a similar behaviour, being attractive in the SO and repulsive in the TO 
channels. However, these contributions computed with 
the  M3Y-P5 force show a rather strong repulsive character in both odd channels. 
In the TE channel the contributions computed with the three interactions have a similar 
behaviour, being attractive upto $k_f\sim$ 2 $fm^{-1}$ and thereafter they become strongly 
repulsive. In the SE channel, the contributions computed with the three forces show the same 
behaviour upto a Fermi-momentum corresponding to normal NM density. Thereafter 
the attraction decreases and becomes repulsive at relatively small $k_f$ value in case of 
SEI and at a larger $k_f$ value for M3Y-P5. On the contrary, the attractive contribution 
of D1S in the SE channel increases when the Fermi momentum $k_f$ increases.   

The density dependence of $isospin$, $spin$ and $spin-isospin$ symmetry 
energies computed with Eqns.~(\ref{eq16}) using the SEI are displayed in the three 
panels of Fig.7.  The corresponding results obtained with the Gogny D1S \cite{blaizot95}, 
D1N \cite{chappert08} and D1M \cite{goriely09} forces are also shown in the 
respective panels. The $isospin$ symmetry energy $E_{\tau}$, in case of the SEI, has a strong 
repulsive density dependence in comparison with the considered Gogny forces 
and does not predict a preferred PNM than SNM at any density.  
In the case of the $spin$ symmetry energy $E_{\sigma}$, however, the Gogny forces 
show a very strong density dependence as compared to the behaviour exhibit by the SEI. 
This is mainly due to the $x_3$ = 1 value in the Gogny forces, which makes the 
contribution of the density dependent term of the spin symmetry energy highly repulsive. 
In the spin channel although the SEI has a relatively soft behaviour, as the D1M Gogny force 
in isospin channel, it does not exhibit instability related to the transition 
of unpolarized to polarized PNM. A value of $E_{\sigma}(\rho_0)$=26.55 MeV 
is obtained in comparison to 27.57 MeV (D1), 29.13 MeV (D1S), 22.69 MeV (D1N), 
28.73 MeV (D1M) and 41.01 MeV (M3Y-P5).  
The $spin-isospin$ symmetry energy $E_{\tau \sigma}$ shows a similar trend 
either computed with the SEI or with the Gogny forces, except D1. It reaches 
a maximum value and decreases becoming zero at certain densities and thereafter 
remains negative. However, the Gogny D1 interaction shows extra stability in the 
$spin-isospin$ channel. The density at which the curve of SEI crosses the $x$-axis 
is larger than the crossing density in the case of the three curves corresponding 
to Gogny D1S, D1N and D1M. A value of $E_{\tau \sigma} (\rho_0)$=23.09 MeV is 
obtained as compared to 30.13 MeV (D1), 29.13 MeV (D1S), 22.69 MeV (D1N), 28.73 MeV 
(D1M) and 41.01 MeV (M3Y-P5). In Fig. 7(a), the symmetry energy, calculated as the 
difference of the energy per particle in PNM and SNM, $E_s (\rho)=[e^N (\rho)-e(\rho)]$ is 
also shown for comparison with the corresponding results of $E_{\tau} (\rho)$ obtained 
from the Taylor series expansion of the energy density in Eqn.(\ref{eq16}).
In NM- calculations, the former definition of the symmetric energy is widely 
used. It is exact 
at the two extremes of $isospin$ asymmetry. The comparison between the two, 
$E_{s}$ and $E_{\tau}$, is excellent over the whole range of density. The 
small 
difference is attributed to
the higher order contributions of the Taylor series expansion which have been 
neglected. 

\begin{figure}
\vspace{0.6cm}
\begin{center}
\includegraphics[width=1.0\columnwidth,angle=0]{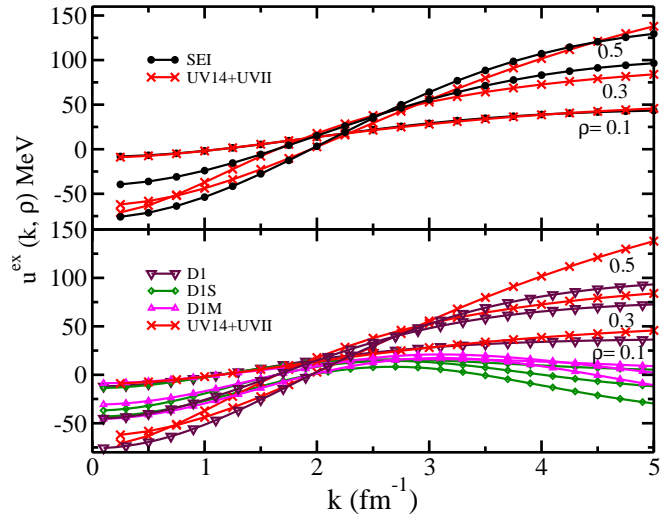}
\caption{(Color Online) The momentum dependent part of the mean field 
$u^{ex}(k,\rho)$ as a function of momentum $k$ for the SEI at three different 
densities, $\rho$=0.1, 0.3, and 0.5 $fm^{-3}$, are compared with the predictions 
of the realistic interaction UV14+UVII \cite{Wiringa} in the upper panel and 
the same comparison for the Gogny D1 \cite{gogny80}, D1S \cite{blaizot95} and 
D1M \cite{goriely09} forces in lower panel.
}
\end{center}

\label{Fig.1}
\end{figure}

\begin{figure}
\vspace{0.6cm}
\begin{center}
\includegraphics[width=1.0\columnwidth]{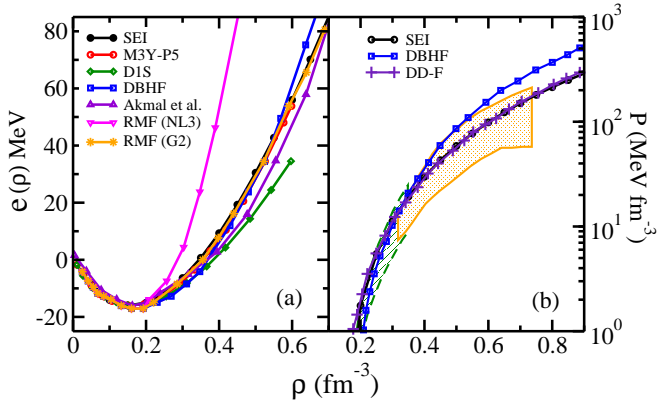}
\caption{(Color Online)(a) Energy per particle in SNM as a function of 
density for 
the SEI compared with D1S \cite{blaizot95}, M3Y-P5 \cite{nakada08}, 
DBHF \cite{samma20}, NL3\cite{patra01}, G2 \cite{patra01} and 
realistic calculation \cite{akmal01}. (b) Pressure as a function 
of density in SNM for SEI compared with DBHF \cite{ebhf04,ebhf05}, 
DD-F \cite{klahn06}, HIC \cite{Daneilewwicz02} and $k^+$ production 
data \cite{lynch09}. See text for details.
}
\end{center}
\label{Fig.2}
\end{figure}

\begin{figure}
\vspace{0.6cm}
\begin{center}
\includegraphics[width=1.0\columnwidth]{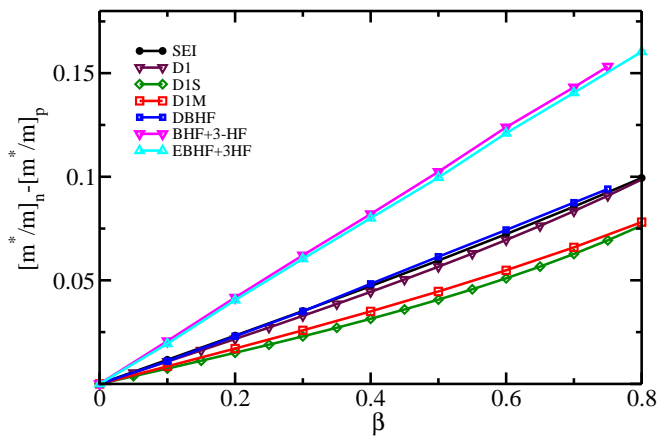}
\caption{(Color Online) The neutron and proton effective mass difference 
in ANM at normal density $\rho_0$ as 
a function of $isospin$ asymmetry $\beta$ for SEI compared with DBHF 
\cite{samma10}, 
BHF+3-BF, EBHF+3-BF \cite{ebhf04,ebhf05,bhf21} and 
Gogny D1 \cite{gogny80}, D1S \cite{blaizot95} and D1M \cite{goriely09} sets.  
}
\end{center}
\label{Fig.3}
\end{figure}

\begin{table*}
\caption{The nine parameters of ANM for the simple effective interaction
(SEI). The connection of the new parameters with the parameters of SEI
is in Eqn. (3).
}
\renewcommand{\tabcolsep}{0.25cm}
\renewcommand{\arraystretch}{1.5}
\begin{tabular}{|c|c|c|c|c|c|c|c|c|c|c|c|}
\hline
\hline
$\gamma$ & $b$    & $\alpha$ & $\varepsilon_{ex}^l$ & $\varepsilon_{ex}^{ul}$ & $\varepsilon_{\gamma}^l$ &
$\varepsilon_{\gamma}^{ul}$ & $\varepsilon_{0}^l$& $\varepsilon_{0}^{ul}$ \\
& $fm$ & $fm$ & $MeV$ & $MeV$ & $MeV$ & $MeV$ & $MeV$ & $MeV$ \\
\hline
$\frac{1}{2}$& 0.5914 & 0.7596 & -94.76 & -125.95 & 77.507 & 97.247 & -78.78 & -111.69 \\
\hline
\multicolumn{9}{|c|}{Nuclear matter properties at saturation condition} \\
\hline
\multicolumn{2}{|c|}{$\rho_0$ ($fm^{-3}$)} & \multicolumn{2}{c|}{$e (\rho_0) $ (MeV)}
& \multicolumn{1}{c|}{$K (\rho_0)$ (MeV)} & \multicolumn{1}{c|}{$\frac{m^*}{m}(\rho_0,k_{f_0})$}
& \multicolumn{1}{c|}{$E_s (\rho_0)$ (MeV)} & \multicolumn{2}{c|}{$L (\rho_0)$ (MeV)} \\
\hline
\multicolumn{2}{|c|}{0.157} & \multicolumn{2}{c|}{-16.0} & \multicolumn{1}{c|}{245}
& \multicolumn{1}{c|}{0.709} & \multicolumn{1}{c|}{35.0} & \multicolumn{2}{c|}{76.26} \\
\hline
\hline
\end{tabular}
\end{table*}

\begin{figure}
\vspace{0.6cm}
\begin{center}
\includegraphics[width=0.9\columnwidth]{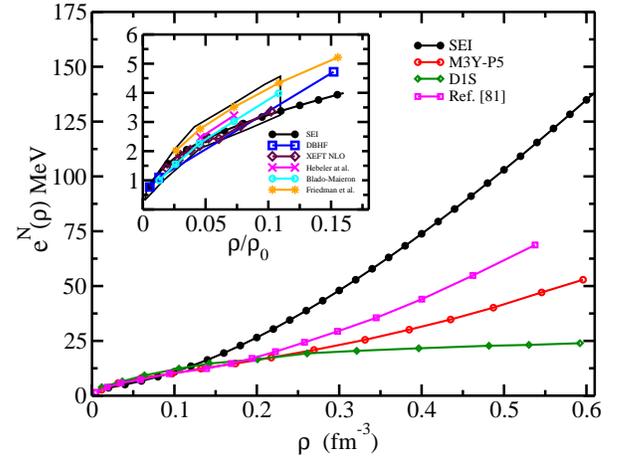}
\caption{(Color Online) The neutron matter EOS for SEI compared with 
D1S \cite{blaizot95}, 
M3Y-P5 \cite{nakada08} and Realistic interaction \cite{pandaripande81}. 
The behaviour in the low density region, $\rho/{\rho_0} < 0.12$, is 
shown in the inset figure, where the Monte Carlo simulation results and 
results of microscopic calculations are compared (see text for details).  
}
\end{center}
\label{Fig.4}
\end{figure}

\begin{figure}
\vspace{0.6cm}
\begin{center}
\includegraphics[width=0.9\columnwidth]{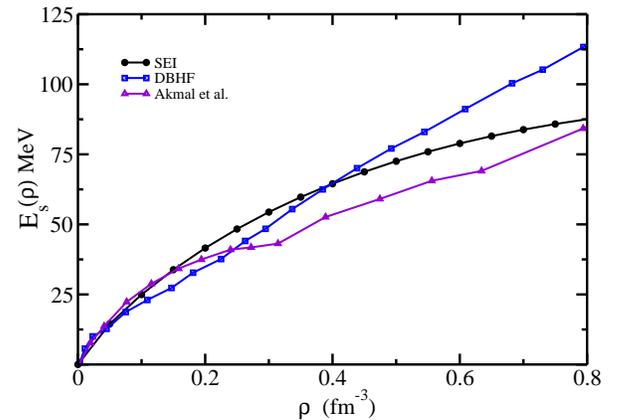}
\caption{(Color Online) Density dependence of the symmetry energy for SEI 
compared with realistic interaction A18+$\delta v$+UIX$^*$ \cite{akmal01} 
and DBHF \cite{ebhf04,ebhf05}.
}
\end{center}
\label{Fig.5}
\end{figure}

\begin{figure}
\vspace{0.6cm}
\begin{center}
\includegraphics[width=0.9\columnwidth]{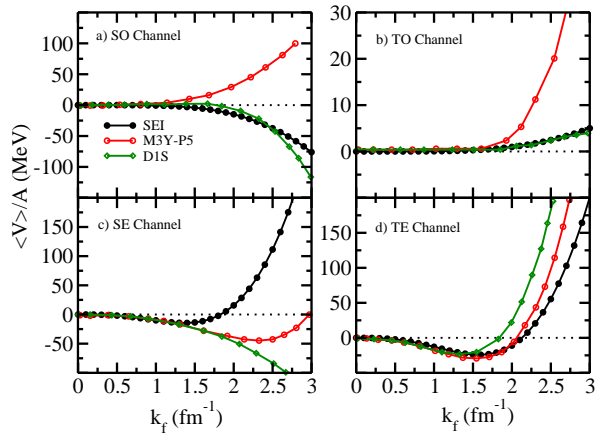}
\caption{(Color Online) Contribution of the SO, TO, SE, and TE channels 
to potential energy per paricle in spin saturated SNM. The predictions 
of Gogny (D1S) \cite{blaizot95} and M3Y-P5 \cite{nakada08} are also shown 
for comparison.
}
\end{center}
\label{Fig.6}
\end{figure}

\begin{figure}
\vspace{0.6cm}
\begin{center}
\includegraphics[width=0.9\columnwidth]{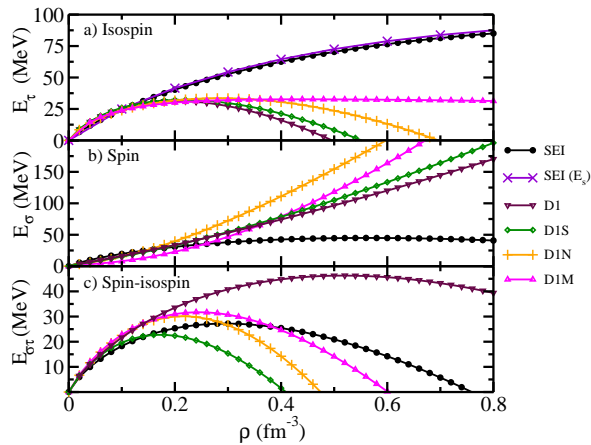}
\caption{(Color Online) The $isospin$, $spin$ and $spin-isospin$ symmetry 
energies are shown as functions of density. The corresponding results for 
Gogny (D1) \cite{gogny80}, (D1S) \cite{blaizot95}, (D1N) \cite{chappert08} 
and (D1M) \cite{goriely09} sets are also shown in the respective figures. 
In (a) the result of $E_s(\rho)$ denoted by SEI($E_s)$ is shown for comparison.
}
\end{center}
\label{Fig.7}
\end{figure}

\section{Finite Nuclei}

As we mentioned at the Introduction, our aim in this Section is to explore 
the ability of the SEI in describing ground-state properties of finite nuclei. 
In Section II we have shown 
that information of NM allows to determine nine of the eleven parameters of 
the SEI. Keeping these nine parameters fixed, we will determine the two 
remaining open parameters, namely $t_0$ and $x_0$, as well as the strength 
of the spin-orbit interaction, $W_0$, from experimental data of some magic nuclei.
Once the SEI is fully determined, we will perform some additional calculations
to check the predictions of our model in other finite nuclei.  We will restrict 
this preliminary study to spherical nuclei only and leave for a future work the 
investigation of the deformation properties of the SEI. For open shell nuclei we
have to add pairing correlations. As we will explain below with some detail,
we use the BCS approach together with a zero-range density-dependent pairing 
interaction that was devised to simulate the behaviour of the Gogny D1 pairing 
gap in neutron matter \cite{garrido99}.

\subsection{The quasilocal Density Functional Theory}  

To deal with finite nuclei, we apply the KS \cite{kohn65} 
method within the framework of the non-local DFT \cite{vinas03}.
As it is proved in this reference, the Lieb theorem \cite{lieb83}, 
which establish the many-to-one mapping of a $A$-particle Slater determinant 
wavefunctions $\Psi_0$ onto the local particle density $\rho({\bf r})$, allows to write 
the energy density functional in the non-local case as $\varepsilon[\rho_0] = 
\varepsilon_0[\rho_0] + E_{RC}[\rho]$. Here $\rho_0$ is the Slater determinant DM
with occupation numbers either 1 or 0 and  $\varepsilon_0$ has the form of the HF energy 
obtained with an effective Hamiltonian $\tilde{H}$. The remaining part, $E_{RC}[\rho]$, 
that is a functional of the local density only, is the residual correlation energy. 
This contribution accounts for the difference between the exact energy functional, 
constructed with the true microscopic Hamiltonian $H$, and the HF energy provided by the 
reference Hamiltonian  $\tilde{H}$. In Ref.~\cite{vinas03} it is also shown that another 
reduction can be performed by mapping the Slater DM onto a set $\rho_{QL}$ of local particle, 
kinetic energy and spin densities for neutron and proton, 
i.e. $\rho^{QL} \equiv \left( \rho_n, \rho_p, \tau_n, \tau_p, {\bf J_n}, {\bf J_p} \right)$.
The particle, kinetic energy and spin local densities entering in the set $\rho^{QL}$
are obtained from the single-particle orbitals $\phi_i$ that define the Slater
determinant $\Psi_0$ as
\begin{eqnarray}
\rho_q ({\bf r})= \sum_{i=1}^{A_q} \sum_{\sigma} \vert \phi_i({\bf r}, \sigma, q)\vert^2,
\end{eqnarray}
\begin{eqnarray}
\tau_q ({\bf r})= \sum_{i=1}^{A_q} \sum_{\sigma} \vert{\bf \triangledown}\phi_i
({\bf r}, \sigma, q)\vert^2,
\label{eq19}
\end{eqnarray}
and
\begin{eqnarray}
J_q ({\bf r})=i \sum_{i=1}^{A_q} \sum_{\sigma,\sigma'} \phi_i^* ({\bf r}, \sigma, q)
\left[({\bf \sigma})_{\sigma,\sigma'} \times{\bf \triangledown}\right] \phi_i
({\bf r}, \sigma, q).
\end{eqnarray}
respectively.

Using this reduction, it is possible to write finally the energy density functional 
in a quasilocal form as $\varepsilon[\rho^{QL}] = \varepsilon_0[\rho^{QL}] + E_{RC}[\rho]$. 
By applying the variational principle to the functional $\varepsilon[\rho^{QL}]$ and
using as functional variables the single-particle orbitals $\phi$ and $\phi^*$, 
one obtains the corresponding KS equations. We shall point out that within 
this quasilocal theory, one would be able to compute the exact ground-state energy and
the exact local particle densities if the exact density functional $\varepsilon[\rho^{QL}]$
is known. However, the kinetic energy densities, $\tau_n$ and $\tau_p$, as well as the spin 
densities, 
${\bf J_n}$ and  ${\bf J_p}$, correspond to uncorrelated system  and therefore do not 
concide with the exact densities within this approach. As explained in \cite{vinas03}, there 
is some freedom
in the choice of the effective Hamiltonian $\tilde{H}$. Therefore, we can choose $\tilde{H}$ as 
an $A$-particle effective interaction of the type
$\tilde{H} = T + \sum_{i\ne j} \hat{v}_{ij}^{NN} + \sum_{i\ne j} \hat{v}_{ij}^{Coul}$.
The nucleon-nucleon interaction $\hat{v}_{ij}^{NN}$ contains the contriutions of the
effective force that we take in this work as the density-independent finite-range part 
of the SEI in Eqn. (\ref{eq1}), and the spin-orbit contribution chosen in the form used 
in the Skyrme and Gogny
interactions: $v_{i,j}^{SO}=iW_0\left({\bf \sigma_i}+{\bf \sigma_j}\right)
\left[{\bf k'\times\delta(r_i,r_j)k}\right]$. From this effective Hamiltonian $\tilde{H}$ 
we obtain the quasilocal energy functional as:
\begin{eqnarray}
\varepsilon_0\left[\rho^{QL}\right]=\int{\mathcal H}_0d^3R,
\label{eq12}
\end{eqnarray}
where the energy density ${\cal H}_0$ reads
\begin{eqnarray}
{\mathcal H}_0&=&\frac{\hslash^2}{2m}\left(\tau_n+\tau_p\right)+
{\mathcal H}_{d}^{Nucl}+{\mathcal H}_{exch}^{Nucl} \nonumber \\ 
 && + {\mathcal H}^{SO} +{\mathcal H}^{Coul}.
\end{eqnarray}
The kinetic energy part corresponds to the non-interacting contribution 
given by Eqn. (\ref{eq19}). The Coulomb energy is taken in the usual way 
as the direct term plus the exchange contribution computed at Slater level 
using the point proton density:
\begin{eqnarray}
{\mathcal H}_{Coul}({\bf r_1})= \frac{1}{2}\int\frac{\rho_p({\bf r_2})}{\vert {\bf r_1-r_2}\vert}d^3r_2
-\frac{3}{4}\left(\frac{3}{\pi}\right)^{1/3}\rho_p^{4/3}({\bf r_1}).
\end{eqnarray}
The spin-orbit energy density, computed using the aforementioned
zero-range force, becomes:
\begin{eqnarray}
{\mathcal H^{SO}}({\bf R}) = -\frac{1}{2}W_0\left[\rho({\bf R}){\bf \nabla J} 
+ \rho_n({\bf R}){\bf \nabla J_n} + \rho_p({\bf R}){\bf \nabla J_p} \right].
\label{eq23}
\end{eqnarray}
The direct contribution to the nuclear energy ${\cal H}_d^{Nucl}$ coming 
from finite range part of the SEI is given by  
\begin{widetext}
\begin{eqnarray}
{\mathcal H}_{d}^{Nucl}&=&\frac{1}{2}\int d^3r_2\left[\left(W+\frac{B}{2}\right) 
\rho({\bf r_1})\rho({\bf r_2}) -\left(H+\frac{M}{2}\right)\left[\rho_n ({\bf r_1})\rho_p({\bf r_2})
+\rho_p({\bf r_1})\rho_n({\bf r_2})\right]f(\vert{\bf r_1-r_2}\vert)\right].
\end{eqnarray}
All these local contributions to the energy density ${\cal H}_0$ constitute the 
so-called Hartree part of the functional. Up to this point, we have developed the {\it exact} theory.
In the next step we shall make some approximations, similar to those used in 
Refs.~\cite{hoffma01,negele72,campi}. To compute the quasilocal energy density corresponding 
to the exchange terms of the nucleon-nucleon force, SEI in this case, we use the Extended
Thomas-Fermi (ETF) expansion of the DM up to $\hbar^2$ order which is widely  discussed 
in Ref.~\cite{vinas00}. For spin-saturated nuclei the exchange nuclear energy density 
splits into two part 
\begin{eqnarray}
{\mathcal H}_{exch}^{Nucl}={\mathcal H}_{exch,0}^{Nucl}+{\mathcal H}_{exch,2}^{Nucl}.
\label{eq17}
\end{eqnarray}
The first term corresponds to the zeroth order of the $\hslash$-expansion 
(Slater approximation for the density matrix) given by
\begin{eqnarray}
{\mathcal H}_{exch,0}^{Nucl}=\int d^3r f(r)\left[
\frac{1}{2}\left(M+\frac{H}{2}-B-\frac{W}{2}\right)\sum_{q=n,p}
\left(\rho_q({\bf R})\frac{3j_1(k_qr)}{(k_q r)}\right)^2+\left(M+\frac{H}{2}\right)
\rho_n({\bf R})\frac{3j_1(k_n r)}{(k_n r)}\rho_p({\bf R})\frac{3j_1(k_p r)}{(k_p r)}
\right], \nonumber \\
\label{eq18}
\end{eqnarray}
where, ${\bf r=r_1-r_2}$ and ${\bf R=\frac{r_1+r_2}{2}}$ are the relative and 
center of mass co-ordinates, respectively.
In Eq.(\ref{eq18}) $k_q({\bf R})=[3 \pi^2 \rho_q({\bf R})]^{1/3}$ and $j_1(x)$
is the spherical Bessel function. The second term of Eqn.(\ref{eq17}) 
is the $\hslash^2-$contribution to the exchange energy which reads \cite{vinas00,vinas03}
\begin{eqnarray}
{\mathcal H}_{exch,2}^{Nucl}=\sum_{q=n,p} \frac{\hslash^2}{2m} \left[\left(g_q-1\right)
\left(\tau_q-\frac{3}{5}k_q^2\rho_q-\frac{1}{4}\nabla^2\rho_q\right)
+k_qg'_q\left(\frac{1}{27}\frac{(\nabla\rho_q)^2}{\rho_q}-\frac{1}{36}\nabla^2\rho_q\right)\right].
\label{eq20}
\end{eqnarray}
Here $g_q=g_q ({\bf R},k_q)$ and $g'_q=(\partial g_q({\bf R},k_q)/\partial k)_{k=k_q}$.  
The function $g_q=g_q ({\bf R},k)$ is the inverse of the position and momentum 
dependent effective mass given by
\begin{eqnarray}
g_q ({\bf R},k)= 1+ \frac{m}{\hslash^2k}\frac{\partial V_{exch,q}^{Nucl} ({\bf R},k)}
{\partial k}.
\label{eq28}
\end{eqnarray}
In this equation $V_{exch,q}^{Nucl}$ is the Wigner transform of the exchange 
potential, 
\begin{eqnarray}
V_{exch,q}^{Nucl}({\bf R},k)=\int d^3r e^{i{\bf k\cdot r}}f(r) 
\left[\left(M+\frac{H}{2}-B-\frac{W}{2}\right)
\rho_q({\bf R})\frac{3j_1(k_qr)}{(k_qr)}
+\left(M+\frac{H}{2}\right)\rho_{q'}({\bf R})\frac{3j_1(k_{q'}r)}{(k_{q'}r)}\right],
\end{eqnarray}
where $q=n,p$ and $q'=p,n$. Notice that in the ANM limit $V_{exch,q}^{Nucl}({\bf R},k)$ is just 
the exchange part of the mean field $u^{q}({\bf k},\rho_q,\rho_{q'})$ in Eqn. (\ref{eq6}) and 
therefore, in this limit, the inverse effective mass $g_q=g_q ({\bf R},k_q)$ reduces to 
Eqn. (\ref{eq11}).

It is worth noting that within the semiclassical ETF expansion of the DM the kinetic energy is a 
functional of the local density only. However, it was found in \cite{vinas00} that the use of the 
quantal kinetic energy density (\ref{eq19}) in the $\hbar^2$-contribution to the exchange energy 
(\ref{eq20}), significantly improves the agreement with the corrersponding full HF calculation.
Therefore, we will use this ansatz here as it was done in previous works \cite{vinas00,
vinas03,krewald06}. It is also important to note that we have repalced the {\it exact}
quasilocal functional $\varepsilon_0\left[\rho^{QL}\right]$ by an approximated one calculated 
with the ETF prescription. The difference between them gives a very small contribution that cannot 
be completely included in the residual correlation energy because the difference depends on 
$\rho^{QL}$ (through $\tau_n$, $\tau_p$, ${\bf J_n}$ and ${\bf J_p}$) while $E_{RC}$ depends  
on $\rho$ only. Notice, however, that this small difference, due to the localization of the exchange
energy, persists if one uses another DM expansions as the Negele-Vautherin \cite{negele72} or 
Campi-Bouyssy \cite{campi} ones.     

Let us now discuss the residual correlation energy 
$E_{RC}$ part of the total energy density. The SEI, as it happens 
with another finite-range forces as the Gogny or M3Y ones, can be split into a density-independent and 
a density-dependent parts. Therefore, a reasonable ansatz, in the spirit of the DFT, is to
take the residual correlation energy $E_{RC}$ as the HF contribution provided by the
 density-dependent part of the interaction in ANM in a local density approximation:   
\begin{eqnarray}
E_{RC} &=& \frac{t_0}{4}\int\left[(1-x_0)\left[\rho_n^2({\bf R})+\rho_p^2({\bf R})\right]
+(4+2x_0)\rho_n({\bf R})\rho_p({\bf R})\right] d^3R\nonumber \\
&&+\frac{t_3}{24}\int\left[(1-x_3)\left[\rho_n^2({\bf R})+\rho_p^2({\bf R})\right]
+(4+2x_3)\rho_n({\bf R})\rho_p({\bf R})\right]\left(\frac{\rho({\bf R})}
{1+b\rho({\bf R})}\right)^{\gamma}d^3R.
\end{eqnarray} 
As mentioned before, the variational principle applied to the full functional 
$\varepsilon[\rho^{QL}] = \varepsilon_0[\rho^{QL}] + E_{RC}[\rho]$
allows to obtain the following set of KS single particle equations:
\begin{eqnarray}
h_q\phi_i={\mathcal E}_i\phi_i,
\label{eq26}
\end{eqnarray}
where, 
\begin{eqnarray}
h_q=-{\bf \nabla}\frac{\hslash^2}{2m^*_q({\bf R})}{\bf \nabla}+ U_q({\bf R})-i{\bf W}_q({\bf R})
\cdot\left[{\bf \nabla}\times{\bf \sigma}\right],
\label{eq27}
\end{eqnarray}
and 
\begin{eqnarray}
\frac{\hslash^2}{2m^*_q({\bf R})}=\frac{\partial {\mathcal E}^{QL}}{\partial \tau_q({\bf R})},\quad
U_q({\bf R})=\frac{\partial {\mathcal E}^{QL}}{\partial \rho_q({\bf R})},\quad
{\bf W}_q({\bf R})=\frac{\partial {\mathcal E}^{QL}}{\partial {\bf J}_q({\bf R})},
\label{eq28}
\end{eqnarray}
\end{widetext}
that are formally similar to the equations of motion obtained with zero-range Skyrme forces.
The quasilocal DFT presented in this work is, actually, very similar to the one used
by Hoffman and Lenske in Ref.~\cite{hoffman98}. The main difference is that the localization 
of the exchange energy is performed here using the ETF expansion of the DM while in 
Ref.~\cite{hoffman98} the Campi-Bouyssy expansion of the DM is used. As it is pointed out
in this reference \cite{hoffman98}, the DM expansions retain, to certain extension, the 
non-local effects of the underlying interaction. In the case of the ETF expansion of the 
DM these effects are collected  in the effective mass and its derivative respect to the 
momentum which appear in Eqn.~\ref{eq20}.

In Ref.~\cite{vinas03} we have chosen the parameters of $\hat{v}_{ij}^{NN}$ in $\tilde{H}$ 
and the ones of $E_{RC}$ to be equal to the parameters of the Gogny D1S force. In this way 
we check to which extend  the quasilocal KS-DFT approach is able to reproduce the full HF 
results with this interaction. We have performed the same comparison with the D1N and D1M
forces. It is found that the difference between the HF and the quasilocal KS-DFT energy 
per nucleon in magic nuclei is always smaller than 
0.01 MeV in $^{48}$Ca, $^{90}$Zr and $^{208}$Pb for all the Gogny forces considered.
 For light nuclei the differences in the binding energy per nucleon are a little 
bit larger but smaller than 0.09 MeV per nucleon in $^{16}$O and than 0.05 MeV per nucleon 
in $^{40}$Ca for these Gogny forces (See also Table I of \cite{vinas00}).
These results show that the quasilocal energy density functional obtained with 
the KS-DFT formalism explained here describes ground-state properties of finite nuclei 
fairly well. This fact motivates us to use directly the quasilocal energy density 
functional $\varepsilon[\rho^{QL}]$ obtained with the SEI fitting its undetermined 
parameters, $t_0$ and $x_0$ of the $E_{RC}$ and the strength of the spin-orbit interaction 
$W_0$, to experiemntal data. 

\begin{table*}
\caption{The parameters of the simple effective interaction (SEI) and spin-orbit strength $W_0$. 
}
\renewcommand{\tabcolsep}{0.25cm}
\renewcommand{\arraystretch}{1.5}
\begin{tabular}{|c|c|c|c|c|c|c|c|c|c|c|c|}
\hline
\hline
$\gamma$ & $b$    & $t_0$     & $x_0$ & $t_3$ &$x_3$     & W     & B     & H     & M    & $\alpha$ & $W_0$ \\
& $fm^3$ & $MeVfm^3$ &       & $MeVfm^{3(\gamma+1)}$& &$MeV$ & $MeV$ & $MeV$ & $MeV$& $fm$     & $MeV$ \\
\hline
$\frac{1}{2}$& 0.5914 & 437.0 & 0.6 & 9955.2 & -0.118 & -589.09 & 130.36& -272.42 & -192.16 & 0.7596 & 115.0 \\
\hline
\hline
\end{tabular}
\end{table*}

\begin{table*}
\caption{The calculated binding energy per particle BE/A, charge radius 
$r_{ch}$, the root mean square radii for neutron $r_n$ and proton $r_p$ are 
compared with the Gogny (D1S) \cite{blaizot95}, M3Y-P5 \cite{nakada08}, 
RMF (NL3$^*$) \cite{nl3im} and experimental data \cite{audi03,angeli04}. The
energy is in MeV and radius in fm.
}
\renewcommand{\tabcolsep}{0.45cm}
\renewcommand{\arraystretch}{1.5}
\begin{tabular}{cccccccccccccccc}
\hline
\hline
Nucleus & Force & BE/A & $r_{ch}$ & $ r_{n}$ & $r_p$  \\
\hline
\hline
$^{16}$O   & SEI        & 7.976 & 2.764 & 2.622 & 2.646  \\
           & Gogny (D1S)& 8.099 & 2.783 & 2.645 & 2.666 & &\\
           & M3Y-P5     & 7.88 &       &       &       & & \\ 
           & RMF (NL3$^*$)  & 8.007 & 2.735 & 2.465 & 2.615 \\
           & Expt.      & 7.976 & 2.730 & &  \\
\hline
$^{40}$Ca   & SEI        & 8.551 & 3.484 & 3.346 & 3.391 \\
            & Gogny (D1S)& 8.616 & 3.501 & 3.365 & 3.408 \\
            & M3Y-P5     & 8.38 &       &       &       \\ 
            & RMF (NL3$^*$)  & 8.539 & 3.470 & 3.237 & 3.377 \\
            & Expt.      & 8.551 & 3.485 & & \\
\hline
$^{48}$Ca   & SEI        & 8.680 & 3.510 & 3.597 & 3.418 \\
            & Gogny (D1S)& 8.681 & 3.534 & 3.583 & 3.442 \\
            & M3Y-P5     & 8.63 &       &       &       \\
            & RMF (NL3$^*$)  & 8.617 & 3.470 & 3.517 & 3.377 \\
            & Expt.      & 8.666 & 3.484 & &  \\
\hline
$^{90}$Zr   & SEI        & 8.708 & 4.275 & 4.285 & 4.199 \\
            & Gogny (D1S)& 8.729 & 4.285 & 4.267 & 4.210 \\
            & M3Y-P5     & 8.66 &       &       &       \\
            & RMF (NL3$^*$)  & 8.693 & 4.263 & 4.227 & 4.187 \\
            & Expt.      & 8.710 & 4.272 & &  \\
\hline
$^{208}$Pb  & SEI        & 7.867 & 5.498  & 5.643  & 5.437 \\
            & Gogny (D1S)& 7.879 & 5.494 & 5.569 & 5.435 \\
            & M3Y-P5     & 7.85 &  &  &  \\
            & RMF (NL3)  & 7.876 & 5.508 & 5.680 & 5.450 \\
            & Expt.      & 7.867 & 5.505 & &  \\
\hline
\hline
\end{tabular}
\end{table*}

\begin{figure}
\vspace{0.75cm}
\includegraphics[width=1.\columnwidth,angle=0]{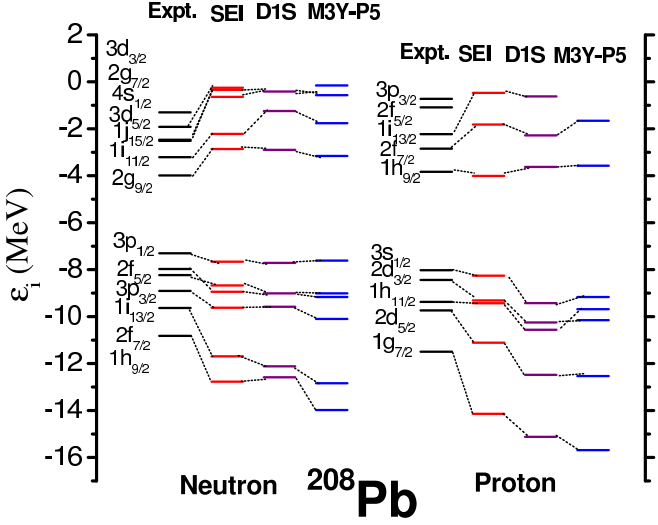}
\caption{The calculated single particle energy levels for $^{208}$Pb are
compared with the Gogny (D1S) \cite{blaizot95}, M3Y-P5 \cite{nakada08}  
and the experimental data \cite{audi95,firestone96}.
}
\label{Fig8}
\end{figure}

\begin{table}[ht]
\caption{The {\it rms} of $S_{2N}$ and $S_{N}$ deviations for SEI compared with
the corresponding results obtained in other HFB calculations.} 
\centering 
\begin{tabular}{c c c c c } 
\hline\hline 
Z or N  & $S_{2N}/S_{N}$  & $S_{2N}/S_{N}$ & $S_{2N}/S_{N}$  \\ 
[0.5ex] 
Chain & SEI & Ref.~\cite{bulgac03} & Ref.~\cite{goriely02} \\ 
\hline 
Z= 20 & 0.85/0.62 & 0.82/0.76 & 1.02/0.92 \\ 
Z= 50 & 0.55/0.62 & 0.29/0.21 & 0.43/0.35 \\
Z= 82 & 0.46/0.82 & 0.23/0.37 & 0.58/0.53 \\
N= 50 & 0.22/0.24 & 0.37/0.26 & 0.41/0.23 \\
N= 82 & 0.21/0.42 & 0.43/0.31 & 0.50/0.56 \\  
N= 126 & 0.67/0.51 & 0.42/0.23 & 0.88/0.52 \\ [1ex] 
\hline 
\end{tabular}
\end{table}

\begin{table}[ht]
\caption{The {\it rms} of $S_{2N}$  deviations for NL3, SLy4, SkM* and D1S } 
\centering 
\begin{tabular}{c c c c c c } 
\hline\hline 
Z or N  & $S_{2N}$  & $S_{2N}$ & $S_{2N}$ & $S_{2N}$  \\ 
Chain & NL3 \cite{nl399} & SLy4 \cite{doba} & SkM* \cite{doba} & D1S \cite{hilaire} \\ 
\hline 
Z= 20 & 0.99 & 0.58 & 1.90 & 0.94 \\ 
Z= 50 & 0.99 & 0.89 & 1.48 & 0.90 \\
Z= 82 & 1.09 & 1.24 & 1.25 & 0.98 \\
N= 50 & 1.05 & 0.53 & 1.03 & 0.93 \\
N= 82 & 1.03 & 0.29 & 1.87 & 0.47 \\  
N= 126 & 1.56 & 0.72 & 2.16 & 0.91  \\ [1ex] 
\hline 
\end{tabular}
\end{table}

\begin{figure}
\vspace{0.75cm}
\begin{center}
\includegraphics[width=0.9\columnwidth]{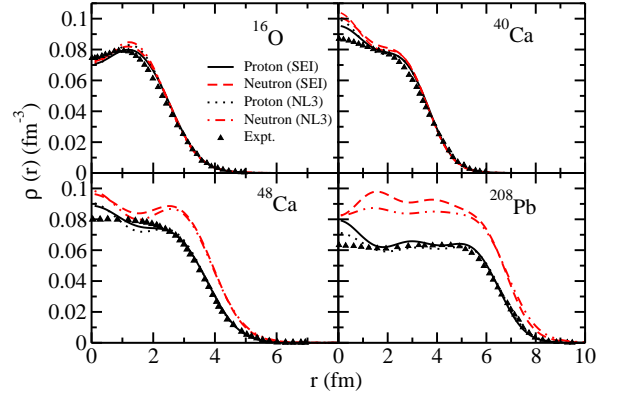}
\caption{The density distribution of protons and neutrons for
$^{16}$O, $^{40}$Ca, $^{48}$Ca and $^{208}$Pb
are compared with the results of NL3 \cite{lala97} and
experimental results of charge distributions \cite{vries87}.
}
\label{Fig7}
\end{center}
\end{figure}

\begin{figure}
\vspace{0.6cm}
\begin{center}
\includegraphics[width=1.0\columnwidth,angle=0]{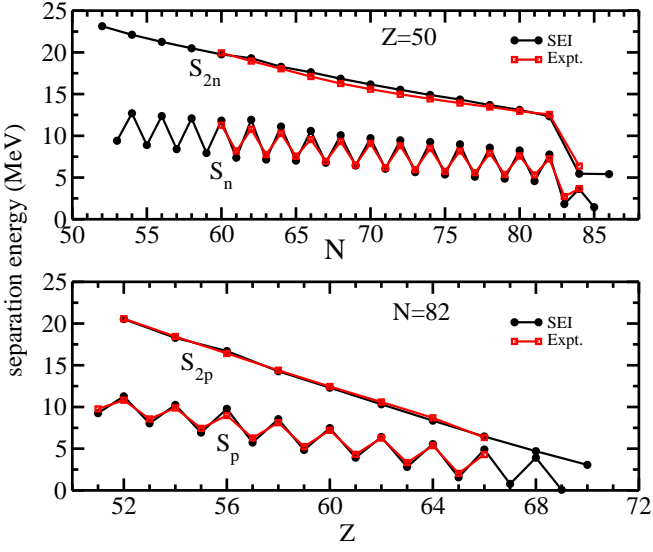}
\caption{(Color Online) The two neutron (proton) and one neutron (proton)
separation energies of Sn isotopes (N=82 isotones) as a function of the 
neutron (proton) number in the upper (lower) panel 
in comparison with the experimental data \cite{audi03}. 
}
\label{Fig6a}
\end{center}
\end{figure}

\begin{figure}
\vspace{0.6cm}
\begin{center}
\includegraphics[width=0.9\columnwidth]{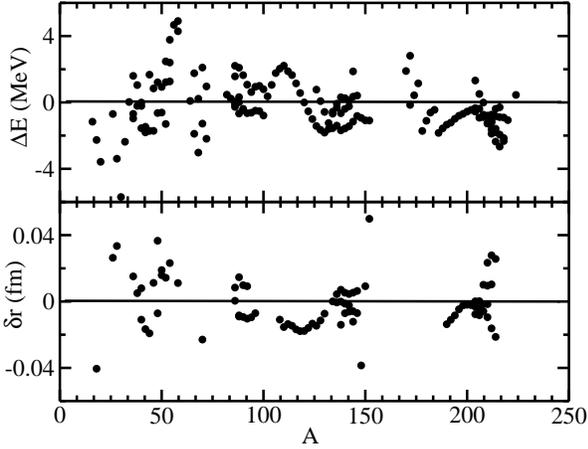}
\caption{The deviation in binding energy $\triangle E$ (upper panel) 
and charge radius $\delta r_{ch}$ (lower panel) of 161 spherical 
even-even nuclei as function of mass number $A$.
The experimental data for binding energies and charge radii
are taken from Refs. \cite{audi03,angeli04}.
}
\label{Fig5}
\end{center}
\end{figure}

\begin{figure}
\vspace{0.6cm}
\begin{center}
\includegraphics[width=0.75\columnwidth,angle=-90]{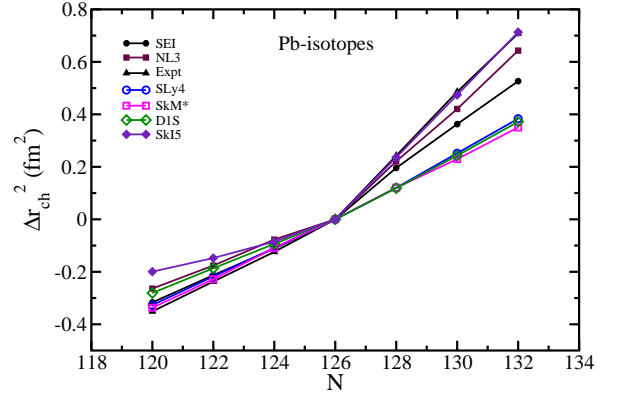}
\caption{(Color Online) 
Isotopic shift
[$\Delta r_{ch}^2=r_{ch}^2(^APb)-r_{ch}^2(^{208}Pb)$]
computed with the SEI energy density functional 
experimental \cite{angeli04}, results of NL3 \cite{nl399}, 
D1S \cite{hilaire}, SLy4, SkM* \cite{doba}, SkI5 \cite{reinhard95}. 
}
\label{Fig6c}
\end{center}
\end{figure}

\begin{figure}
\vspace{0.6cm}
\begin{center}
\includegraphics[width=0.75\columnwidth,angle=-90]{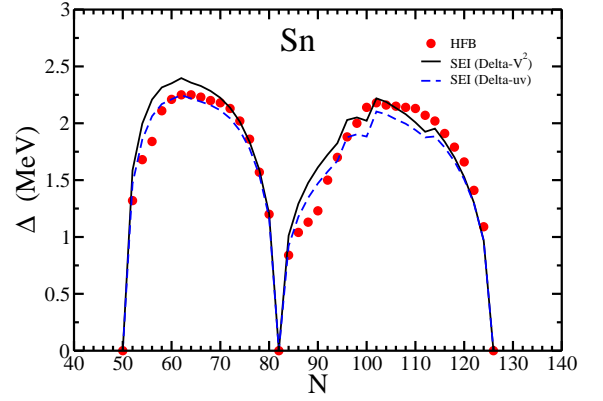}
\caption{(Color Online) The average pairing gap from SEI for
constant gap and constant strength approximation compared with
the HFB prediction \cite{dobac96}.
}
\label{Fig6}
\end{center}
\end{figure}

\begin{figure}
\vspace{0.6cm}
\begin{center}
\includegraphics[width=1.0\columnwidth,angle=0]{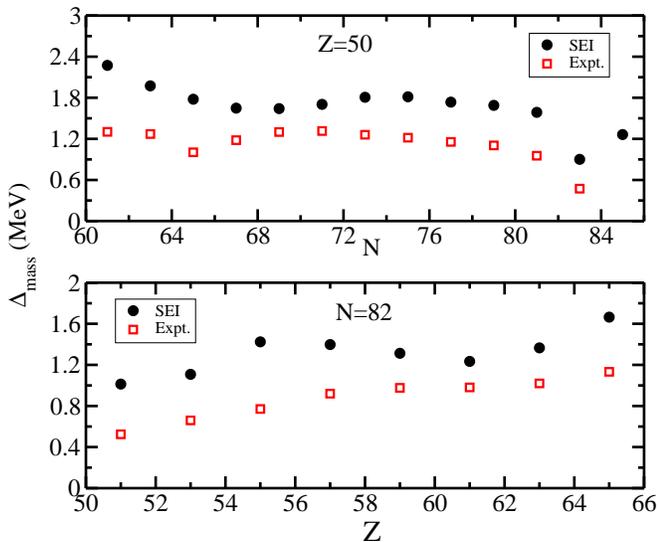}
\caption{(Color Online) The odd-even mass staggering of Sn isotopes computed
with the SEI energy density functional compares with the experimental 
values (upper panel). The same for isotones of N=82 (lower panel).
}
\label{Fig6b}
\end{center}
\end{figure}

Certainly, pairing correlations play an important role for open shell nuclei. 
To deal with such nuclei, it is mandatory to include the pairing correlations.
The first formal generalization of the Hohenberg-Kohn theorem to paired systems 
was performed in Ref. \cite{oliveira88} for superconductors. Some modifications 
of this approach were developed later in Refs.~\cite{lathi04,luders05}. More recently,
in Ref.~\cite{krewald06}, an extension of the DFT including pairing correlations 
without formal violation of the particle-number conservation and  
the quasilocal reduction of this non-local theory are discussed with detail.
The equations of motion associated to this energy density functional including 
pairing correlations have the same form as the HFB equations. 
However, for pairing calculations of nuclei not too far from the
$\beta$-stability line, the simpler BCS approach can be sufficient for describing
their ground-state energies \cite{bender00}. For these nuclei, the Fermi level 
lies appreciably below zero and, consequently, the levels around it, which mainly
contribute to pairing correlations, are also well bound avoiding the 
problems of the standard BCS approach near the drip lines \cite{dobac84,dobac96}. 
For practical BCS calculations, one extracts from the
energy density functional a part which depends only on the normal density
and that describes the nuclei without pairing effects. The remaining part   
contains the contributions arising from the pairing correlations. In the BCS 
approach the motion equations, in the case of spherical symmetry, reduces to a 
set of mean-field single-particle equations, that in the quasilocal approach 
are given by Eq.~(\ref{eq26}), and another set of gap equations
\begin{eqnarray}
\Delta_{i} = - \sum_k \frac{2j_k + 1}{4 \pi} V^{pp}_{ik}\frac{\Delta_k}{2 E_k},\nonumber\\
E_{i} = \sqrt {\left(\lambda - {\mathcal E}_{i}\right)^{2} + {\Delta_i}^{2}}.
\label{eq34}
\end{eqnarray}
In the gap equations (Eqn. (\ref{eq34})), $E_i$ are the quasiparticle energies with 
${\mathcal E}_{i}$ the single-particle energies, $\Delta_i$ the state
dependent gaps and $\lambda$ the chemical potential to ensure the right number 
of particles. In this equation $V^{pp}_{ik}$ are the reduced matrix 
elements of the effective interaction in the particle-particle channel. Eqns.~(\ref{eq26})
and (\ref{eq34}) are coupled among them and have to be solved self-consistenly. 
In Refs.~\cite{krewald06,vinas07} the ability of this quasilocal extension of
the DFT including pairing correlations at BCS level to reproduce full HFB binding 
energies and gaps was checked. The calculations were performed for several Sn and Pb isotopes 
lying in the stability valley using the Gogny D1 and D1S forces in both, particle-hole
and particle-particle, channels. In these test calculations the differences between 
the full HFB and the quasilocal approach plus BCS are less than 0.2\%, which are similar 
to the values found for magic nuclei \cite{vinas03}. 

In this paper we will use this formalism for the study of the binding energies of some 
open shell spherical nuclei. To this end we will use the mean field provided by the SEI. 
For a sake of simplicity we chose as pairing interaction a density-dependent zero-range 
force of the type proposed by Bertsch and Esbensen \cite{bertsch91}: 
\begin{eqnarray}
v({\bf r_1},{\bf r_2}) = V_0\bigg[1 - \eta \bigg(\frac{\rho(\frac{{\bf r_1}+{\bf r_2}}{2})}
{\rho_0}\bigg)^{\alpha} \bigg]\delta({\bf r_1}-{\bf r_2}).                                     
\label{eq35}
\end{eqnarray}
This effective pairing force is extensively used in nuclear structure calculations 
\cite{sandu04,sandu05,krewald06,baldo08,junli08,grill11,pastore11}.
The numerical values of the parameters
$V_0$=-481 MeV $fm^3$, $\eta$=0.45, $\alpha$=0.47 and $\rho_0$=0.16 $fm^{-3}$
are taken from Ref.~\cite{garrido99}. These parameters, together with a cutoff energy 
$\epsilon_C$=60 MeV from the bottom of the single-particle potential, 
were fitted in \cite{garrido99} to reproduce the gap values of 
the Gogny D1 force in neutron matter. We use here an improved BCS approach \cite{patra01}
where the resonant levels \cite{dobac96} are simulated very efficiently by quasibound 
single-particle energy levels retained by the centrifugal (neutrons) or centrifugal plus 
Coulomb (protons) barriers. 
The two-body center of mass correction has been taken into account self 
consistently by using a pocket formula based on the harmonic oscillator and 
derived in Ref.~\cite{butler84}. 

\subsection{Results of finite nuclei}

In order to determine the three parameters, $t_0$, $x_0$ and $W_0$, we proceed 
as follows. Keeping the nine parameters determined in ANM (see Table I) fixed, 
(i) We reproduce the experimental binding energy of $^{40}$Ca by
adjusting $t_0$, (ii) The spin-orbit strength parameter $W_0$ is 
adjusted to reproduce the experimental splitting of the neutron and proton $1p$ levels 
in $^{16}$O (iii) The parameter $x_0$ is obtained by fitting the binding 
energy and charge radius of $^{208}$Pb. The steps (i)-(iii) are repeated till 
the self-consistency of the parameters $t_0$, $x_0$ and $W_0$ is achieved.
The value of the eleven parameters of the SEI entering in Eqn.~(\ref{eq1}) 
as well as the strength of the spin-orbit force are given in Table II.

The binding energy per particle (BE/A) and the proton ($r_p$), neutron ($r_n$) 
and charge ($r_{ch}=\sqrt{r_p^2+0.64}$ $fm$) {\it rms} radii of the 
standard magic nuclei computed with the SEI quasilocal energy density functional 
are displayed in Table III. These results are compared with the corresponding values obtained 
using the Gogny (D1S) \cite{gogny84,blaizot95} and the  M3Y-P5 
\cite{nakada08} forces and  the RMF (NL3$^*$) parametrization \cite{nl3im} 
as well as with the experimental data (binding energies from \cite{audi03} and 
charge radii from \cite{angeli04}) that are also displayed in Table III. 
It is observed that the BE/A obtained with the SEI functional matches nicely with the 
experimental data for these magic nuclei . 
In case of radii, our results are similar to those provided 
by the Gogny and the RMF calculations. The neutron skin thickness, defined as the 
difference between the neutron and proton {\it rms}, i.e. $=r_n-r_p$, in $^{208}$Pb 
predicted by our calculation using the SEI functional is 0.21 $fm$. This value is 
in agreement with the experimental result obtained in Ref. \cite{znihiro10} 
using proton-nucleus scattering.
 

Experimental information about single-particle energies of even-even nuclei can 
be obtained from the low-lying excited states of the adjacent odd nuclei by adding
or picking up a single nucleon. The ordering of the particle and hole levels, for 
both neutrons and protons, predicted by the SEI functional is in agreement with the 
experiment \cite{audi95,firestone96} except for the neutron $2f_{5/2}$ level which 
lies below the $3p_{1/2}$ one. This is, however, a relatively common fact in many 
mean field models, as for example the well reputed NL3 \cite{bender03}. The level 
spacing predicted by the SEI functional is larger than the experimental one, but 
similar to the spacing obtained with the D1S and M3Y-P5 forces. The spin-orbit 
splittings of the single-particle levels $3p$ and $2f$ of neutrons and $2d$ of 
protons obtained with our functional overestimate the experimental values
as it also happens with the D1S and M3Y-P5 predictions.

The radial dependence of the density distributions for protons 
and neutrons of the nuclei $^{16}$O, $^{40}$Ca,
$^{48}$Ca and $^{208}$Pb are displayed in the four panels of Fig.~\ref{Fig7}. 
The neutron and proton densities for the same nuclei obtained using the RMF 
(NL3) \cite{lala97} alongwith the experimental data for charge distributions, 
taken from Ref.~\cite{vries87}, are also given for
comparison in the same figure. For $^{16}$O, $^{40}$Ca and $^{48}$Ca the
agreement between our densities and the ones predicted by NL3 model is quite
good. The density distributions of $^{208}$Pb computed with SEI and NL3 differ
more between them, mainly in the bulk region. The proton densities predicted
by the SEI functional agree reasonably well with the experimental charge distributions
at the surface. In the interior of the nucleus the quantal oscillations shown
by our proton densities are, in general, well averaged.

From Table III it is seen that the SEI functional predicts ground-state energies
and radii that accurately reproduce the experimental values. In this respect, 
the following comments are in order. To determine the nine parameters of ANM,
we have considered the empirical values of three NM parameters, namely $\rho_0$,
$e(\rho_0)$ and $E_s(\rho_0)$. The widely accepted ranges of these parameters
are $\rho_0$=0.17$\pm$0.03 $fm^{-3}$, $e(\rho_0)$=-16$\pm$0.02 MeV and
$E_s(\rho_0)$=30-35 MeV. We have examined the variations of these parameters 
within their ranges and have found that small variations in the values of $\rho_0$,
$e(\rho_0)$ and $E_s(\rho_0)$ have a large influence on the predictions of
binding energies and radii in finite nuclei, although the changes in the
NM predictions are not of much significance. It is also found that $\rho_0$
critically depends on the value of $\gamma$ that determines the stiffness
of EOS in SNM. For a given value of $\gamma$, there is a critical value
of $\rho_0$ for which the predictions of binding energies and radii will
have minimal deviation from experimental results. Thus for the EOS considered
in this work corresponding to $\gamma$=1/2, the best results in finite nuclei
are found for $\rho_0$=0.157 $fm^{-3}$, $e(\rho_0)$= -16 MeV and $E_s(\rho_0)$
=35 MeV.

We have performed further investigations to check the ability of our proposed
SEI functional to describe ground-state properties of spherical nuclei.
In this exploratory calculation pairing correlations are treated at BCS by
the reasons pointed out before. Notice, however, that the BCS approach for
pairing has been used in several well known mass tables as the ones computed
with NL3 \cite{nl399} and HFBCS-1 \cite{goriely01}.
In the upper panel of Fig.~\ref{Fig6a} we display the two-neutron and the 
one-neutron separation energies for Sn isotopes and in the lower panel of the
same figure the two-proton and one-proton separation energies for the N=82 
isotonic chain. To obtain the one-neutron and one-proton separation
energies we have to compute odd-even or even odd-nuclei. To deal with these nuclei
we perform, as in Ref.~\cite{nakada08}, spherical blocking calculations which neglect 
part of the core polaritazion effect in the odd nuclei \cite{bender00}. In these
calculations one has to block several single-particle states around the Fermi
energy and search for the configuration giving the lowest energy. From Fig.~\ref{Fig6a},
we see that the SEI predictions match fairly well with the experimental values for 
both chains. To be more quantitative, we have also computed the two- and one-nucleon 
separation energies in other isotopic and isotonic chains with magic proton and neutron 
numbers, respectively. In Table IV we show the {\it rms} deviation of these separation 
energies with respect to the experimental values \cite{audi03} for the $Z$=20, 50, 82 
isotopic and $N$=50, 82, 126 isotonic chains. Our predictions are compared with the HFB
results of Refs.~\cite{bulgac03} and \cite{goriely02} (see the former reference for more 
details). A similar comparison between our results and the ones provided by 
the SLy4 and Gogny D1S forces and the NL3 parameter set, but only for two-nucleon separation 
energies is given in Table V. From Table IV, it can be seen that our results are not so good 
than those of \cite{bulgac03} but similar to the ones of \cite{goriely02}. Notice, however, 
that in Refs.~\cite{bulgac03} and \cite{goriely02} the strength of the pairing force is fitted to
some experimental data while in our case the pairing interaction is taken from the
literature without any refit of their parameters to experimental data.
The quality of the our predictions for two-nucleon separation energies also compare 
very satisfactorily with the results obtained from HFB calculations with 
SLy4 and SkM$^*$ \cite{doba} and D1S \cite{hilaire} forces and the ones calculated with 
the NL3 parameter set \cite{nl399}. 

We have performed more tests to check the ability of the SEI energy density for 
describing ground-state properties of finite nuclei. To this end,  we have 
computed the binding energy and charge radii of 161 even-even spherical 
nuclei between $^{16}$Ne and $^{224}$U. The difference between the theoretical 
prediction and the experimental value for binding energy $\triangle E$ 
and charge radius $\delta r_{ch}$ are shown in the upper and lower panel of 
Fig. 11, respectively. In general, the deviations in binding energies and charge radii 
for the $161$-spherical nuclei are broadly reproduced within $\pm 2$ MeV and 
$\pm 0.02$ $fm$, respectively, baring few exceptions. 
The overall {\it rms} deviation in energy and charge radius of these nuclei are 
$(\Delta E)_{rms}$ =1.5402 MeV and $(\Delta R)_{rms}$ = 0.0152 $fm$, respectively. These 
values are a little bit smaller than the corresponding rms deviations obtained 
with well calibrated effective interactions as D1S, SLy4, NL3 and BCP for the same set of 
nuclei (see Table 3 of Ref.~\cite{baldo08}). We have also estimated 303 binding energies
and 111 charge radii of odd spherical nuclei, for which the experimental values are known.
For these nuclei we find $(\Delta E)_{rms}$ =1.6501 MeV and $(\Delta R)_{rms}$ = 
0.0198 $fm$, respectively.

Using our SEI, we have also computed the isotopic shift of charge radii, 
defined as $\Delta r_{ch}^2=r_{ch}^2(^APb)-r_{ch}^2(^{208}Pb)$, in
Pb-isotopes. Analyzing the $\Delta r_{ch}^2$ values obtained
with the SEI, one can see in Fig.~\ref{Fig6c} that they lie quite close to the experimental
values \cite{angeli04} and that the kink at $A$=208 is reasonably well
reproduced. It is known that Skyrme and Gogny forces use isospin independent spin-orbit 
interactions those are unable to reproduce the experimental kink exhibit by the 
isotopic shift of the charge radius in Pb isotopes at the double magic $^{208}$Pb 
\cite{reinhard95,sharma95}. This mismatch can be cured by introducing an isovector 
spin-orbit contribution as it is the case of the SkI family of Skyrme forces 
\cite{reinhard95} or using non-linear RMF parametrizations \cite{sharma93}. However, 
the SEI functional incorporates a spin-orbit contribution that does not contain isovector 
part (see Eqn.(\ref{eq23})) but clearly shows up the kink without imposing it as a constraint 
in the fitting procedure of the functional. A possible explantion is the following.
In a recent paper \cite{arnau13} it is claimed that the development of the kink in the 
isotopic shift of the charge radius of $^{208}$Pb is largely determined by the occupation of the 
$1i_{11/2}$ neutron level. This orbital has a principal quantum number $n$=1 and overlaps strongly 
with the majority of the proton orbitals. This produces a relatively large pulling of the proton 
orbitals by the neutron ones via the symmetry energy increasing the proton radius, and, therefore, 
developing the kink at $A=208$. From Fig.8 we can see that the SEI functional predicts that the 
$1i_{11/2}$ lies close to the $2g_{9/2}$ one and the same happens for heavier Pb isotopes. 
Therefore, one can expect a relatively large occupation of $1i_{11/2}$ state. This fact together 
with the relatively high symmetry energy of the SEI seems to be the reason for the appearence 
of the kink exhibit by the SEI results in Fig.12.     

We shall now discuss in some detail the pairing properties of our SEI energy 
density functional. This functional includes a pairing contribution coming from 
a zero-range density-dependent force that simulates the Gogny interaction in the 
particle-particle channel. As far as the Gogny pairing force gives a good description of 
finite nuclei combined with a reasonable mean field, not only provided by 
the Gogny interaction but also by Skyrme forces and RMF parametrizations
\cite{nl3im}, we are, actually, checking the single-particle energies
obtained with the SEI energy density functional. 
In Fig.~\ref{Fig6} we display the average 
gaps along the whole Sn isotopic chain obtained with our formalism 
compared with the values predicted by a full HFB calculation with the 
Gogny D1S force in both, particle-hole and particle-particle, channels. 
From this figure it can be seen that the averaged gaps, defined as 
$\bar{\Delta}_{v^2}= \sum_n v_n^2 \Delta_n/\sum_n v_n^2$ and
$\bar{\Delta}_{uv}= \sum_n u_n v_n \Delta_n/\sum_n u_n v_n$
nicely reproduce the HFB values in this isotopic chain
taken of Ref.~\cite{dobac96}.
This result suggests that our model, with a mean field part whose 
parameters are fitted to ANM and magic nuclei together with a realistic 
pairing force taken from the literature, is, in principle,  well suited for describing 
open-shell nuclei at mean field level with a quality similar to that found
using well known effective interactions or RMF paramtrizations.

A possible way of estimating pairing correlations from an experimental 
point of view is through the so-called odd-even mass staggering (see 
e.g. Ref.~\cite{bender00}). In Fig.~\ref{Fig6b} we display the odd-even 
mass staggering using a three-point formula \cite{bender00}
\begin{eqnarray}
\Delta_{mass} &=&
-\frac{1}{2}\bigg[E(N+1)-2E(N)-E(N-1)\bigg]
\end{eqnarray}
for the Sn isotopic chain (upper panel) and for the $N$=82 isotonic chain
computed with the SEI energy functional in comparison with the experimental 
values. 
Our theoretical prediction of the mass staggering slightly
overestimate the experiemental values in both, Sn isotopic chaim and $N$=82
isotonic chain. Therefore, some comments are in order. 
On the one hand, the odd-even mass staggering of experimental masses is not a 
pure measure of the pairing correlations \cite{bender00}. It also contains mean 
field contributions related to the rotational and time-reversal symmetries breaking 
not accounted by the simple blocking approach.
On the other hand, the fact that the 
theoretical odd-even mass staggering overestimates the corresponding
experimental values may be not so dramatic. It is known that 
when pairing is computed using the Gogny interaction, the 
average gaps are larger than the mass gap \cite{gogny80}. However, 
as discussed in this reference, effects beyond mean-field, such as 
quasiparticle vibration coupling, are expected to reduce the average 
gaps. 
Finally to point out that in this exploratory calculation no attempt has been done
in order to adjust simultaneously pairing and mean field. Therefore, it seems reasonably  
that a slightly different protocol in the fitting procedure of the
parameters $t_0$ and $x_0$, for example including some open shell in it, would allow to
find an odd-even mass staggering in better agreement with the experimental data.
From all the discussion developed along this section, we believe that
the reported calculations show up clearly that the proposed energy 
density functional based on the SEI is a realiable tool for dealing, at 
least in spherical nuclei, with  quality similar to that found using 
well calibrated Gogny or Skyrme forces or succesful RMF paraametrizations 
in the same scenario.

\section{Summary and conclusions}

We have proposed a simple effective interaction aimed to describe accurately the 
main trends of microscopic calculations in nuclear and neutron matter and, at 
the same time, to reproduce the ground state properties of finite nuclei with a 
quality similar to the one obtained using successful effective interactions of 
Skyrme, Gogny and M3Y type or relativistic mean field model. Our interaction contains a single 
Gaussian form factor for describing the finite range part of the force plus a 
zero-range part that includes a density dependent term to simulate the effect 
due to three-body forces. Our effective interaction depends on eleven parameters, nine of 
them fitted to nuclear and neutron matter data. 
The SNM is fully determined by six parameters, which are three strength combinations, 
the range of the force, the exponent $\gamma$ and the factor $b$. Two of these 
six parameters, the strength of the exchange interaction and the range of the 
force, completely determine the momentum dependence of the mean field in SNM. 
These two parameters are constrained by using the optical potential data 
which imposes that the calculated mean field in normal SNM vanishes at a 
kinetic energy $300$ MeV of the incident nucleon. 
The exponent $\gamma$ of the density dependent term is chosen amongst the 
values that gives the pressure-density relation in agreement with the experimental 
results of HIC and $K^+$ production. The parameter $b$ is fixed to avoid the
supraluminous behaviour in NM. The two remaining strength combinations in SNM are 
determined from the saturation conditions.
In our model, the range of the force is chosen to be the same for interactions
between pairs of like and unlike nucleons. Under this restriction, the splitting of the 
exchange strength parameter in ANM is decided from the condition that the entropy in PNM
does not exceed the one in SNM.
The splitting of the two other strengths into like and unlike channels in ANM are decided 
from the value of the symmetry energy and by imposing that the nucleonic
part of the energy density in charge neutral $\beta$-stable matter be a maximum. 
With the nine parameters of our effective interaction determined in this way,  
the corresponding EOS and mean field follow, quite closely, the general 
trends shown up by sophisticated microscopic calculations.

To describe finite nuclei we use this interaction to build up an energy density functional 
in the framework of the quasilocal Density Functional Theory. In particular we use the 
semiclassical $\hbar^2$- expansion of the density matrix to deal with the 
exchange contribution. This procedure allows to write the single particle equations 
in finite nuclei in a similar form as in the case of the Skyrme-Hartree-Fock equations. 
We determine the open parameters of our interaction $t_0$ and $x_0$, and the strength of the 
spin-obit force $W_0$ by a simple fit to experimental data of three closed shell nuclei 
$^{16}$O, $^{40}$Ca and $^{208}$Pb. With this choice the binding energies and charge 
radii of the standard magic nuclei are accurately reproduced. With our simplified interaction 
fully determined, we explore its predictive power for describing open-shell 
spherical nuclei. To this end, we add to our functional the contribution of a zero-range
pairing interaction, taken from the literature, which reproduces the Gogny neutron gaps in 
PNM. In this exploratory calculation we use the BCS approximation instead of HFB theory for
dealing with open shell nuclei and the simplified spherical blocking approach to estimate 
the ground-state properties of odd nuclei. It is important to note that within this framework, 
the results for open-shell nuclei are predictions of our model as far as no parameter has been 
fitted to open-shell nuclei. We have computed the energy of $161$ and charge  
radii of $88$ spherical even-even nuclei in the mass region $A=$ 16-224. Our calculation 
reproduce the experimental values fairly well. The {\it rms} deviations of the binding energies i
and radii of these nuclei with respect to the experimental values are $1.5402$ MeV and $0.0152$ 
$fm$, respectively, which are slightly better than the same results for the set of nuclei 
computed using standard interactions such as Gogny D1S, BCP, NL3 and SLy4. We have also 
calculated the one- and two-nucleon separation energies along the Ca, Sn and Pb isotopic chains
and $N$=50, 82 and 126 isotonic chains. With our model, the {\it rms} deviation of the theoretical
predictions respect to the experimental data compares favourably with the results obtained
using well calibrated effective interactions of Skyrme, Gogny and relativistic mean field type.
We have also investigated the isotopic shift of the charge-radii in Pb. 
We find, surprisingly, that the kink shown by the experimental charge radii at $A$=208, is
qualitatively reproduced by our model, in spite that our spin-orbit force has no isospin 
dependence, which is needed, in principle, to reproduce the kink with non-relativistic 
mean field models. We have given a possible explanation of this fact based on the small 
gap between the $2g_{9/2}$ and $1i_{11/2}$ neutron levels and the relatively
large symmetry energy predicted by the SEI model.

In this exploratory analysis of finite nuclei properties described with our simplified 
effective interaction we have restricted to spherical nuclei. To improve the treatment 
of pairing correlations by replacing the BCS approach  by a Hartree-Fock-Bogoliubov
calculation, and extend our study to deformed nuclei and some collective excited states
is a necessary task to confirm the success of our simplified effective interaction 
that will be developed in future works.

\section*{Acknowledgments}
The authors are indebted to P. Schuck and L.M. Robledo by useful disucssions.
This work is supported in part by the UGC-DAE Consortium for Scientific Research, 
Kolkata Center, Kolkata, India (Project No. UGC-DAE CRS/KC/CRS/2009/NP06/1354) and 
the work is covered under SAP program of School of Physics, Sambalpur University,
India. One author (TRR) thanks Department d'Estructura i Constitutuents de Materia, 
University de Barcelona, Spain for hospitality during the visit.
B.K.S and X.V. acknowledges the support of the Consolider Ingenio 2010 Programme 
CPAN CSD2007-00042, Grant No. FIS2011-24154 from MICINN and FEDER, and Grant 
No. 2009SGR-1289 from Generalitat de Catalunya. B.K.S also acknowledge the support
Grant No. CPAN10-PD13 from CPAN (Spain).  One of the author MB acknowledges 
the support in part by Council of Scientific $\&$ Industrial Research 
(File No.09/153(0070)/2012-EMR-I).

\end{document}